# Axial and pseudoscalar current correlators and their couplings to η and η' mesons

J. P. Singh[1] and J. Pasupathy[2]
[1]Physics Department, Faculty of Science, The Maharaja Sayajirao University of Baroda, Vadodara-390002, India
[2]Center for High Energy Physics, Indian Institute of Science, Bangalore-560012, India

**Abastract :** Correlators of singlet and octet axial currents, as well as anomaly and pseudoscalar densities have been studied using QCD sum rules. Several of these sum rules are used to determine the couplings $f_\eta^8$, $f_\eta^0$, $f_{\eta'}^8$ and $f_{\eta'}^0$. We find mutually consistent values which are also in agreement with phenomenological values obtained from data on various decay and production rates. While most of the sum rules studied by us are independent of the contributions of direct instantons and screening correction, the singlet-singlet current correlator and the anomaly-anomaly correlator improve by their inclusion.

## I. INTRODUCTION

The determination of the decay constants of the $\eta$ and $\eta'$ mesons for the octet and singlet axial vector currents is of great interest both experimentally and theoretically. The constants are defined by

$$\left\langle 0 \right| J_{\mu 5}^a \left| P(p) \right\rangle = i f_P^a p_\mu , \qquad \text{..... (1)}$$

where the index $a$ = 8,0 denotes the octet and singlet currents respectively. In terms of u, d and s quark fields the currents are defined by

$$J_{\mu 5}^8 = \frac{1}{\sqrt{6}}(\bar{u}\gamma_\mu\gamma_5 u + \bar{d}\gamma_\mu\gamma_5 d - 2\bar{s}\gamma_\mu\gamma_5 s) \qquad \text{.....(2a)}$$

$$J_{\mu 5}^0 = \frac{1}{\sqrt{3}}(\bar{u}\gamma_\mu\gamma_5 u + \bar{d}\gamma_\mu\gamma_5 d + \bar{s}\gamma_\mu\gamma_5 s) . \qquad \text{.....(2b)}$$

The pseudoscalar meson state P of momentum p can be either $\eta$ or $\eta'$. The four couplings $f_\eta^8$, $f_{\eta'}^8$, $f_\eta^0$ and $f_{\eta'}^0$ occur in the determination of a number of production and decay amplitudes involving $\eta$ and $\eta'$. Among the nonet of pseudoscalars $\pi$, K, $\eta$ and $\eta'$, the isosinglets $\eta$ and $\eta'$ are of special interest because of the so-called U(1) problem [1-6] and the presence of an anomaly in the divergence of the axial singlet current. Thus, one has

$$\partial^\mu J_{\mu 5}^0 = \frac{2i}{\sqrt{3}}(m_u \bar{u}\gamma_5 u + m_d \bar{d}\gamma_5 d + m_s \bar{s}\gamma_5 s) - \frac{\sqrt{3}}{4}\frac{\alpha_s}{\pi} G\tilde{G} \qquad \text{......(3)}$$

$$\partial^\mu J_{\mu 5}^8 = \frac{2i}{\sqrt{6}}(m_u \bar{u}\gamma_5 u + m_d \bar{d}\gamma_5 d - 2m_s \bar{s}\gamma_5 s) \qquad \text{......(4)}$$

Where

$$G\tilde{G} = ½\ \varepsilon^{\mu\nu\rho\sigma} G_{\mu\nu}^a G_{\rho\sigma}^a , \ \varepsilon^{0123} = +1 \qquad \text{......(5)}$$

It was shown by 't Hooft [2] that in the presence of instanton configurations in the vacuum, chirality can change in the vacuum and therefore there is no Nambu-Goldstone boson associated with the singlet current. In other words, $\eta'$ is massive even in the chiral limit $m_u = m_d = m_s = 0$. Let us denote by $m_\chi$

$$m_{\eta'}(m_u = m_d = m_s = 0) = m_\chi. \qquad ........(6)$$

It was shown by Witten [3] and Veneziano [5] using $1/N_c$ expansion, where $N_c$ is the number of colors, that

$$m_\chi^2 = 12\,\chi(0)\big|_{GD} / F^2 \qquad ......(7)$$

where $\chi(0)\big|_{GD}$ is the topological susceptibility in gluodynamics (GD), i.e., in SU($N_c$) gauge theory without any quark fields. F is the pion decay constant in the large $N_c$ limit. (In our normalization Eq.(2a), $f_\pi$ = 131 MeV.) According to the counting rules $F^2$ goes as $N_c$ in the large $N_c$ limit, then $m_\chi \to 0$ as $\eta'$ also becomes massless [3]. The anomaly-anomaly correlator is generically defined as

$$\chi(q^2) = i\int d^4x e^{iqx} \langle 0| T\{Q_5(x), Q_5(0)\} |0\rangle , \qquad .....(8)$$

where

$$Q_5(x) = (\alpha_s/8\pi)\; G^a_{\mu\nu}(x)\tilde{G}^{a\mu\nu}(x) . \qquad ....(9)$$

In pure gluodynamics, $\chi(q^2)$ is evaluated in the SU(3) gauge theory while in the real world QCD, quark fields with appropriate masses should be included in evaluating Eq.(8).

Returning to the four constants $f_P^a$ ($a$ = 0,8; P = $\eta$, $\eta'$) defined in Eq. (1), following current literature we write them in the matrix form

$$\begin{pmatrix} f_\eta^8 & f_\eta^0 \\ f_{\eta'}^8 & f_{\eta'}^0 \end{pmatrix} = \begin{pmatrix} f_8\cos\theta_8 & -f_0\sin\theta_0 \\ f_8\sin\theta_8 & f_0\cos\theta_0 \end{pmatrix} \qquad .....(10)$$

in terms of two mixing angles $\theta_8$ and $\theta_0$. Until a decade ago, partly to eliminate one parameter, it was customary to use just one mixing angle with Eq. (10) written as a product of a diagonal matrix of octet and singlet constants and an orthogonal matrix corresponding to the octet singlet mixing in $\eta$ and $\eta'$ states.

Using chiral perturbation theory which includes the U(1) sector, Kaiser and Leutwyler [7-8] derived the relation

$$\sin(\theta_0 - \theta_8) = 2\sqrt{2}\,\left(F_K^2 - F_\pi^2\right)/\left(4F_K^2 - F_\pi^2\right) \qquad .....(11)$$

Here $F_K$ is the kaon decay constant

$$F_K/F_\pi - 1 = 0.22 . \qquad ....(12)$$

Eqs. (11) and (12) clearly show that $\theta_0 \neq \theta_8$ and indeed a more coherent picture of $\eta$ and $\eta'$ decays emerge when one uses two distinct angles $\theta_0$ and $\theta_8$, instead of the earlier approach which assumed $\theta_0 = \theta_8$. During the last decade a number of authors have carried out phenomenological analysis which is summarized in Table III.

A number of theoretical approaches have been used to compute the four constants $f_8$, $f_0$, $\theta_8$ and $\theta_0$. Apart from chiral perturbation theory [7-9], Shore [10] has computed them using the so called generalized Dashen-Gell-Mann-Oakes-Renner (DGMOR) program. A number of theoretical papers based on QCD sum rules have also appeared [11-16].

In an earlier work [17], we had computed the derivative of the topological susceptibility at zero momentum $\chi'(0)$ and determined the mass of $\eta'$ in the chiral limit as well as singlet decay constant in the same limit. For $\chi'(0)$ we obtained a value $\approx 1.82\times10^{-3}$ GeV$^2$. This is close to the value $1.9\times10^{-3}$ GeV$^2$ obtained in Ref.[18] using only the axial vector current sum rules. Further it was used to determine the isosinglet axial vector coupling $\langle p,s|\, \bar{u}\gamma_\mu\gamma_5 u + \bar{d}\gamma_\mu\gamma_5 d \;|p,s\rangle$ which along with $G_A$, and the octet coupling $G_8$ successfully account for the Bjorken sum rule. In a complimentary approach Ioffe and his collaborators [13,14] used the experimental data on Bjorken sum rule to determine $\chi'(0)$ and found

$$\chi'(0) = (2.3\pm0.6)\times10^{-3}\ \text{GeV}^2. \qquad \dots (13)$$

Further, Leutwyler [19] in chiral perturbation theory found that

$$\chi'(0) = \frac{F^2}{4} m_{red}^2 \left[\frac{1}{m_u^2} + \frac{1}{m_d^2} + \frac{1}{m_s^2}\right] + \frac{1}{2} H_0 \qquad \dots (14)$$

where

$$(m_{red})^{-1} = m_u^{-1} + m_d^{-1} + m_s^{-1}, \qquad \dots (15)$$

$H_0$ is a parameter in the effective chiral Lagrangian that describes low energy QCD and is expected to be small. Leutwyler estimates the first term in Eq.(14) which depends only on quark mass ratios and not on their absolute values to be $2.2\times10^{-3}$ GeV$^2$, which is consistent with the determination in Ref.[13,14,17,18 ].

In Ref.[17], we had estimated the mass of $\eta'$ in the chiral limit to be

$$m_{\eta'}(m_q = 0) = 723\ \text{MeV}, \qquad \dots (16a)$$

$$F_0(m_q = 0) = 178\ \text{MeV}. \qquad \dots (16b)$$

Returning to Witten-Veneziano formula Eq.(7), let us note that $\chi(0)\big|_{GD}$ has been determined in the lattice to be [20]:

$$\chi(0)\big|_{GD} = (191\pm5)^4\ \text{MeV}^4. \qquad \dots (17a)$$

On the other hand , we can determine $\chi(0)\big|_{GD}$ using Eq.(7) and Eqs. (16a) and (16b) above, which gives

$$\chi(0)\big|_{GD} = (193\ \text{MeV})^4 \qquad \dots (17b)$$

in excellent agreement with the lattice value Eq. (17a)

The above discussion on $\chi'(0)$ and Eqs.(17a,b) suggests that despite the various approximations involved, QCD sum rule method can be a useful tool to determine the values of $f_8$ , $f_0$ , $\theta_8$ and $\theta_0$, which is the main theme of the current work.

The paper is organized as follows: In the next section we introduce several functions which a priori can be useful to compute the four constants $f_8$ , $f_0$ , $\theta_8$ and $\theta_0$. We briefly discuss the various low energy theorems and briefly discuss the QCD sum rule method and point out that replacement of the correlators by the operator product expansion can violate low energy theorems, and therefore introduce poles at $q^2$=0 while the exact function has none. In Sec. III, we write down the OPE for the various correlators and corresponding sum rules for the various functions of interest. In Sec. IV, we analyze the fits for the sum rules and extract the values of $f_8$ , $f_0$ , $\theta_8$ and $\theta_0$ from the residues from the sum rules that we consider satisfactory. In Sec. V, we discuss the sum rules that do not work well and also briefly comment on the work of other authors. An Appendix gives brief details of low energy theorems and effective chiral Lagrangian.

## II. FORMALISM

Following Ioffe [12-15], we introduce the correlator of axial vector currents

$$\Pi^{ab}_{\mu\nu}(q) = \mathrm{i}\int d^4x e^{iqx} \langle 0|T\ \{J^a_{\mu 5}(x), J^b_{\nu 5}(0)\}\ |0\rangle\ ;\quad (a,\mathrm{b}=8,0). \qquad \text{.....(18)}$$

The general form of the polarization tensor $\Pi^{ab}_{\mu\nu}(q)$ is

$$\Pi^{ab}_{\mu\nu}(q) = -\,\mathrm{P}^{ab}_L(q^2)\,\mathrm{g}_{\mu\nu} + \ \mathrm{P}^{ab}_T(q^2)\,(-\,\mathrm{q}^2\,\mathrm{g}_{\mu\nu} + \mathrm{q}_\mu\,\mathrm{q}_\nu) \qquad \text{....(19)}$$

The functions $\mathrm{P}^{ab}_L(q^2)$ and $\mathrm{P}^{ab}_T(q^2)$ are free from kinematic singularities. On forming the divergence with the momentum, we get

$$\mathrm{q}^\mu\ \Pi^{ab}_{\mu\nu}(q)\,\mathrm{q}^\nu\ = -\,\mathrm{P}^{ab}_L(q^2)\,\mathrm{q}^2 \qquad \text{....(20)}$$

On the other hand from Eq. (18) we have the Ward identity [14]

$$\mathrm{q}^\mu\ \Pi^{00}_{\mu\nu}(q)\,\mathrm{q}^\nu\ = \mathrm{i}12\int d^4x e^{iqx}\langle 0|\mathrm{T}\{\mathrm{Q}_5(\mathrm{x}),\,\mathrm{Q}_5(0)\}\,|0\rangle + \mathrm{i}2\int d^4x e^{iqx}\langle 0|\mathrm{T}\{\mathrm{Q}_5(\mathrm{x}),\,\mathrm{D}(0)\}\,|0\rangle$$

$$+\ \mathrm{i}2\int d^4x e^{iqx}\langle 0|\mathrm{T}\{\mathrm{D}(\mathrm{x}),\,\mathrm{Q}_5(0)\}\,|0\rangle$$

$$+\ \mathrm{i}\,\frac{1}{3}\int d^4x e^{iqx}\langle 0|\mathrm{T}\{\mathrm{D}(\mathrm{x}),\,\mathrm{D}(0)\}\,|0\rangle + \frac{4}{3}\sum_{i=u,d,s} m_i\,\langle o|\bar{q}_i q_i\,|0\rangle\,. \qquad \text{.....(21)}$$

In Eq.(21) we have introduced the notation

$$\mathrm{D}(\mathrm{x}) = 2\mathrm{i}\sum_{i=u,d,s} m_i\bar{q}_i(x)\gamma_5 q_i(x)\,. \qquad \text{...(22)}$$

Following Ioffe we note

$$\lim_{\mathrm{q}\to 0}\ \mathrm{q}^\mu\ \Pi^{00}_{\mu\nu}(q)\,\mathrm{q}^\nu\ = \lim_{\mathrm{q}\to 0} -\,\mathrm{P}^{00}_L(q^2)\,\mathrm{q}^2$$

$$= 0 \qquad \text{....(23)}$$

since the invariant $\mathrm{P}^{00}_L(q^2)$ is regular at $q^2 = 0$. This low energy theorem, namely the vanishing of the left hand side of Eq.(21), has been studied in detail by Ioffe [14 ]. In particular, he noted that the contributions of the Goldstone states, which are linear in quark masses must vanish separately in the right hand side of Eq. (21) for zero momentum. The special nature of the matrix elements of the anomaly $Q_5(x)$ between the vacuum and Goldstone states plays a crucial role. One has the following results

$$\langle 0|Q_5|\pi\rangle = -\frac{1}{2\sqrt{2}}\frac{m_u - m_d}{m_u + m_d}\,f_\pi m_\pi^2 \qquad \text{....(24)}$$

$$\langle 0|Q_5|\eta\rangle = \frac{1}{2}\sqrt{\frac{1}{6}}\,f_\pi m_\eta^2 \qquad \text{...(25)}$$

which shows that the anomaly matrix elements are far from flavor symmetric and linear in quark masses. In Eq.(24)

$$\frac{m_u - m_d}{m_u + m_d} = O(1)$$

and we have the GMOR relation

$$m_\pi^2 = -2(m_u + m_d)\ \langle 0|\bar{q}q|0\rangle / f_\pi^2$$

$$m_\eta^2 = -\frac{8}{3} m_s \left(1 - \frac{1}{4}\frac{m_u + m_d}{m_s}\right) \langle 0|\bar{q}q|0\rangle / f_\pi^2$$

so that the matrix elements (24) and (25) are linear in quark masses. The intermediate states other than the π and η occurring in Eq.(21) have nonzero masses in the chiral limit. We can therefore separately consider terms linear in the light quark masses in analyzing the low energy theorem Eq.(23) which lead Ioffe to obtain the result [14]:

$\chi(0) = m_{red} <0|qq|0>$ + higher order terms in quark masses. …(26)

Let us briefly consider the method of QCD sum rules. Denoting generically

$$F(q^2) = \frac{i}{\pi}\int d^4x e^{iqx} \langle 0|T\{A(x), B(0)\}|0\rangle$$

where A(x) and B(x) are the local fields that connect the vacuum to the hadronic state of interest, one considers the dispersion relation

$$F(q^2) = \frac{1}{\pi}\int \frac{\mathrm{Im} F(s)}{s - q^2} ds + \text{subtractions}$$

and Borel transforms it to obtain

$$\hat{B}\, F(q^2) = \frac{1}{\pi}\int \mathrm{Im} F(s) e^{-s/M^2} ds \qquad \text{….(27)}$$

where the Borel transform is defined by

$$\hat{B}\, F(q^2) = \lim_{-q^2\to\infty,\, n\to\infty} \left[\frac{(-q^2)^{n+1}}{n!}\left(\frac{d}{dq^2}\right)^n F(q^2)\right]_{-q^2/n = M^2}$$

Now the left hand side of Eq.(27) is computed using the operator product expansion while the right hand side is written in the form

$$\frac{1}{\pi}\int \mathrm{Im} F(s) e^{-s/M^2} ds = \lambda_H\, e^{-m_H^2/M^2} + \frac{1}{\pi}\int_{W^2}^{\infty} \mathrm{Im} F(s) e^{-s/M^2} ds \qquad \text{… (27a)}$$

where $\lambda_H$ is the coupling involving the lowest mass state H in the dispersion representation:

$\mathrm{Im}\, F(s) = \pi\, \lambda_H\, \delta\left(s - m_H^2\right)$ + contributions from higher mass states.

This leads to

$$\lambda_H\, e^{-m_H^2/M^2} = \hat{B}\, F(q^2) - \frac{1}{\pi}\int_{W^2}^{\infty} \mathrm{Im} F(s) e^{-s/M^2} ds. \qquad \text{….(27b)}$$

One matches the left-hand side and the right-hand side over some $M^2$ interval to determine $\lambda_H$ and $m_H$. There are several issues to be addressed here: (1) which function $F(q^2)$ should one choose where there is more than one choice – in our case instead of $P_L^{88}$ we could have chosen the function describing the correlator

$$i \int d^4x e^{iqx} \langle 0 | T\{D(x), D(0)\} | 0 \rangle ,$$

(2) what $W^2$ one should choose for the second term in Eq.(27a), and (3) what is the $M^2$ region over which we should match the left hand side and right hand side in Eq.(27b)? These are all related questions. The choice of $F(q^2)$ is dictated by its asymptotic behavior for large $q^2$. If the choice is between, say

$$F_1(q^2) \sim q^4 \ln(-q^2)$$

and

$$F_2(q^2) \sim q^2 \ln(-q^2),$$

$F_2(q^2)$ is to be preferred since higher mass states in Im $F_2(s)$ are less dominant as compared to higher mass states in Im $F_1(s)$. One can at best, make an estimate of the higher mass state contribution by using duality, that is, one equates the sum over excited states by the smeared average as given by the perturbative loop in $F_2(q^2)$. Clearly $W^2$ should be close to the squared mass of the first excited state which one expects to be in the range 2 to 2.5 $GeV^2$. Using a significantly higher value of $W^2$ invalidates Eqs.(27a,b). Similarly the interval in $M^2$ is dictated by the following. In computing $\hat{B}\, F(q^2)$ using OPE, we are usually able to calculate only a small number of higher dimensional operators. The smaller the $M^2$ is the more important are the higher dimensional operators which puts a lower limit on $M^2$, while the larger the $M^2$ is the more important are the excited states in Eqs.(27a,b) which puts an upper limit on $M^2$. This, therefore, determines the $M^2$ interval over which Eqs. (27a,b) can be expected to be valid. The constants $m_H$ and $\lambda_H$ are then obtained by looking for the best fit for Eqs.(27a,b). It is easy to see that if one fits $m_H$ at the experimental mass, this leads to a better determination of the coupling since $m_H$ appears in the exponential. It should be borne in mind that the sum rule results are subject also to the errors in values of the vacuum expectation values for the various condensates.

We shall consider several functions : $P_L^{00}(q^2)$, $P_L^{08}(q^2)$, $P_L^{88}(q^2)$, $\frac{\chi(q^2)}{q^2}$, $\frac{\chi'(q^2)}{q^2}$, $-P_L^{00}(q^2) - 12\frac{\chi(q^2)}{q^2}$ and $S(q^2)$ which are discussed in the Sec.IV. Before that we turn to the OPE for the various T-products that are needed

## III. OPERATOR PRODUCT EXPANSION AND DIRECT INSTANTON CONTRIBUTION

We will be using the following operator product expansion, cf. Refs.[ 11,21,22]

$$i\int d^4x e^{iqx}\left\langle 0\right| T\{Q_5(x), Q_5(0)\}\left|0\right\rangle =$$

$$-(\frac{\alpha_s}{8\pi})^2\frac{2}{\pi^2}q^4\ln(\frac{-q^2}{\mu^2})[1+\frac{\alpha_s}{\pi}(\frac{83}{4}-\frac{9}{4}\ln(-\frac{q^2}{\mu^2}))]$$
$$-\frac{1}{16}\frac{\alpha_s}{\pi}\left\langle 0\left|\frac{\alpha_s}{\pi}G^2\right|0\right\rangle(1-\frac{9}{4}\frac{\alpha_s}{\pi}\ln(\frac{-q^2}{\mu^2}))+\frac{1}{8q^2}\frac{\alpha_s}{\pi}\left\langle 0\left|\frac{\alpha_s}{\pi}g_s G^3\right|0\right\rangle$$
$$-\frac{15}{128}\frac{\pi\alpha_s}{q^4}\left\langle 0\left|\frac{\alpha_s}{\pi}G^2\right|0\right\rangle^2+16(\frac{\alpha_s}{4\pi})^3\sum_{i=u,d,s}m_i\left\langle \overline{q_i}q_i\right\rangle[\ln(-\frac{q^2}{\mu^2})+\frac{1}{2}]$$

$$-\frac{1}{2}\int d\rho\, n(\rho)\rho^4 q^4 K_2^2(Q\rho) + \text{screening correction to the direct instanton} \qquad \ldots(28)$$

The perturbative term above, is taken from Kataev et. al [22]. The so called direct instanton (DI) terms and their screening are described in detail by Forkel[21]. In the constant density or spike approximation

$n(\rho) = n_0\delta(\rho-\rho_c)$

one gets

$$i\int d^4x e^{iqx}\left\langle 0\right| T\{Q_5(x), Q_5(0)\}\left|0\right\rangle|_{DI} \simeq \frac{1}{4}n_0 M^3\rho_c^3\sqrt{\pi}e^{-M^2\rho_c^2}\left(M^2\rho_c^2+\frac{13}{4}+\frac{165}{32}\frac{1}{M^2\rho_c^2}\right) \ldots(29)$$

while in the Gaussian-tail approximation [21]

$$n_G(\rho) = \frac{2^{18}}{3^6\pi^3}\frac{\bar{n}}{\bar{\rho}}\left(\frac{\rho}{\bar{\rho}}\right)^4\exp\left(-\frac{2^6}{3^2\pi}\frac{\rho^2}{\bar{\rho}^2}\right),\ N_f=N_c=3,\ \bar{\rho}\cong 0.6\ \text{GeV}^{-1},\ \bar{n}\cong 7.53\times10^{-4}\ \text{GeV}^4, \ldots(30)$$

$$i\int d^4x e^{iqx}\left\langle 0\right| T\{Q_5(x), Q_5(0)\}\left|0\right\rangle|_{DI} = -\frac{1}{2}\int d\rho\, n_G(\rho)\rho^4 q^4 K_2^2(Q\rho) \qquad \ldots(30a)$$

and the integration has to be performed numerically. Forkel has also extensively described the screening corrections to the above, caused by correlations between instatons, which are very important. We shall return to this point later.

For the crossed correlation between the anomaly and psuedoscalar density, we have

$$i(-2im_s)\int d^4x e^{iqx}\left\langle 0\right|T\{Q_5(x)\ ,\ \bar{s}\gamma_5 s(0)\}\left|0\right\rangle$$

$$=m_s^2 q^2\ln(-\frac{q^2}{\mu^2})\left\{\frac{\gamma}{2}-\frac{7}{4}+\frac{1}{4}\ln(-\frac{q^2}{\mu^2})\right\}$$

$$+\left(\frac{\alpha_s}{\pi}\right)^2 m_s\left\langle\bar{s}s\right\rangle\ln(-\frac{q^2}{\mu^2})-\frac{1}{4}\frac{\alpha_s}{\pi}\left\langle\frac{\alpha_s}{\pi}G^2\right\rangle\frac{m_s^2}{q^2}\ln(-\frac{q^2}{\mu^2})$$

$$+\frac{\alpha_s}{2\pi}m_s\left\langle\bar{s}g_s\sigma.Gs\right\rangle\frac{1}{q^2}. \qquad \ldots(31)$$

The expressions in Eq. (31) is the result of our independent calculations.

For the psuedoscalar density –density correlation we have the OPE [11]

$$i(2i\,m_s)^2\int d^4x e^{iqx}\left\langle 0\right|T\{\ \bar{s}\gamma_5 s(x), \bar{s}\gamma_5 s(0)\ \}\left|0\right\rangle$$

$$= -\frac{3}{2\pi^2}\ \mathrm{m}_s^2 q^2 \left[ \ln\left(-\frac{q^2}{\mu^2}\right) - 2 + \left\{ -\frac{131}{12} + \frac{17}{3}\ln(-\frac{q^2}{\mu^2}) - \frac{11}{3}\ln^2\left(-\frac{q^2}{\mu^2}\right)\right\}\right]$$

$$+ 8\ m_s^3 \left\langle \bar{s}s \right\rangle \frac{1}{q^2} - m_s^2 \left\langle \frac{\alpha_s}{\pi} G^2 \right\rangle \frac{1}{q^2} + 4\, m_s^2 \left\{ \frac{16}{3}\pi^2 \frac{\alpha_s}{\pi} \left\langle \bar{s}s \right\rangle^2 + m_s \left\langle \bar{s} g_s \sigma.Gs \right\rangle \right\} \frac{1}{q^4}$$

…(32)

The last dimension five and six terms above have been computed by us.

## IV. ANALYSIS AND DISCUSSION

Our interest in this work is to determine the couplings listed in Eq.(10) . The seven functions listed in the last paragraph of Sec.II, contain differing combinations of these four couplings. They have differing asymptotic $q^2$ behavior, and differ in the remaining non-perturbative terms given the various vacuum condensates. We fix the η and η′ masses at their experimental values

$m_\eta$ = 0.547 GeV, $m_{\eta'}$ = 0.958 GeV.

For the other quantities needed in the sum rules, we shall use the values (cf. Refs.[17,14])

$\alpha_s$ (1GeV) = 0.5, $a = -(2\pi)^2 \left\langle \bar{q}q \right\rangle$ = 0.55GeV$^3$, b= $\left\langle g_s^2 G^2 \right\rangle$=0.5 GeV$^4$,

$\left\langle \bar{q} g_s \sigma.Gq \right\rangle = m_0^2 \left\langle \bar{q}q \right\rangle$ with $m_0^2$=0.8 GeV$^2$, $\left\langle \bar{s}s \right\rangle$=0.8$\left\langle \bar{u}u \right\rangle$, $\left\langle g_s \frac{\alpha_s}{\pi} G^3 \right\rangle = \frac{\varepsilon}{2} \left\langle \frac{\alpha_s}{\pi} G^2 \right\rangle$ with

$\varepsilon$ = 1.0 GeV$^2$, and $m_s$ = 0.153 GeV. …(33)

We first begin with the octet-octet correlator, $\mathrm{P}_L^{88}(q^2)$ obtaining from the Eq. (20) with both a=b=8. We have from Eq. (4)

$$-\, \mathrm{q}^2\, \mathrm{P}_L^{88}(q^2) = \mathrm{i}\, q^\mu q^\nu \int d^4x e^{iqx} \left\langle 0 \right| T\, \left\{ J_{\mu 5}^8(x), J_{\nu 5}^8(0) \right\} \left| 0 \right\rangle$$

$$= \mathrm{i} \int d^4x e^{iqx} \left\langle 0 \right| T\, \left\{ \partial^\mu J_{\mu 5}^8(x), \partial^\nu J_{\nu 5}^8(0) \right\} \left| 0 \right\rangle + \mathrm{ETCR} \qquad \ldots(34)$$

The Borel transformed sum rule is obtained following the procedure of Eqs. (27), (27a) and (27b). Since we must include η, η′ mixing, the ground state hadrons consist of both η and η′. In Sec.II it was pointed out that in $\mathrm{P}_L^{88}(q^2)$, because of division by $q^2$ in l.h.s. of Eq. (20), suppression of the excited state contributions will lead to a better sum rule than, for example, the one obtained from $S(q^2)$ in Eq. (39) below. However, care is needed. First note that for the the left hand side of Eq.(20) is zero when $q^2$ is zero .This means that the exact function $\mathrm{P}_L^{88}(q^2)$ is regular at $q^2$=0. As noted by Ioffe [14] , the vanishing at $q^2$ = 0 of the right had side of Eq.(34) results from the cancellation of Goldstone state contribution at $q^2$ = 0 and ETCR. As explained in detail in the Appendix, the replacement of the T product

$$T\{D_s(x), D_s(0)\} \qquad \ldots(35)$$

by its operator product expansion valid for large $q^2$ can lead to a violation of the low-energy theorem

$$\lim_{q_\mu \to 0} q^{\mu}\, \Pi^{88}_{\mu\nu}(q)\, q^{\nu} = \lim_{q_\mu \to 0} - q^2\, P^{88}_{L}(q^2) = 0 \qquad \ldots(36)$$

Hence the approximate $P^{88}_{L}(q^2)$ obtained by OPE used in the sum rule introduces a spurious pole in $P^{88}_{L}(q^2)$ at $q^2 = 0$ whose residue we denote by $K^{88}$. With this, the sum rule reads

$$K^{88} + m^2_{\eta'}\left(f^8_{\eta'}\right)^2 \exp(-m^2_{\eta'}/M^2) + m^2_{\eta}\left(f^8_{\eta}\right)^2 \exp(-m^2_{\eta}/M^2)$$

$$= \frac{1}{\pi^2} m_s^2 M^2 \left\{1 + \frac{\alpha_s}{\pi}\left(\frac{17}{3} + 2\gamma - 2\ln\frac{M^2}{\mu^2}\right)\right\} E_0\left(W^2/M^2\right) - \frac{8}{3} m_s \left\langle \bar{s}s\right\rangle$$

$$+ \frac{16}{3} m_s^3 \frac{1}{M^2}\left\langle \bar{s}s\right\rangle - \frac{2}{3} m_s^2 \frac{1}{M^2}\left\langle \frac{\alpha_s}{\pi} G^2\right\rangle - \frac{64}{9}\pi^2 \frac{\alpha_s}{\pi} m_s^2 \frac{1}{M^4}\left\langle \bar{s}s\right\rangle^2$$

$$- \frac{4}{3} m_s^3 \frac{1}{M^4}\left\langle \bar{s} g_s \sigma.G s\right\rangle . \qquad \ldots(37)$$

To extract $\left(f^8_{\eta'}\right)^2$ and $\left(f^8_{\eta}\right)^2$ we need to specify the range of $M^2$ over which the left hand side and the right hand side match and the value of the continuum threshold $W^2$. In this and following sum rules we use the criterion that at the lower end of $M^2$, the contribution of the highest dimensional term to the OPE side be less than 5% and at the higher end of $M^2$, the continuum state contributions be less than 32% of the sum of all terms in the right hand side. In Eq. (37) we use a value $W^2 = 2.3$ GeV$^2$ and the results of fitting Eq.(37) in the range $1.0$ GeV$^2 \le M^2 \le 1.7$ GeV$^2$ are displayed in Fig.1. We find

$K^{88} = 1.097\times10^{-3}$ GeV$^4$, $\left(f^8_{\eta}\right)^2 m^2_{\eta} = 8.20\times10^{-3}$ GeV$^4$, $\left(f^8_{\eta'}\right)^2 m^2_{\eta'} = 3.55\times10^{-3}$ GeV$^4$. …(38)

leading to the values

$f_8 = 176.8$ MeV and $|\theta_8| = 20.6^{\circ}$. …(38a)

Next, we write the pseudoscalar density correlator

$$S(q^2) = i\int d^4x e^{iqx} \left\langle 0\right| T\{m_s\bar{s}(x)\gamma_5 s(x), m_s\bar{s}(0)\gamma_5 s(0)\}\left|0\right\rangle \qquad \ldots(39)$$

We have the sum rule

$$m^4_{\eta'}\left(f^8_{\eta'}\right)^2 \exp(-m^2_{\eta'}/M^2) + m^4_{\eta}\left(f^8_{\eta}\right)^2 \exp(-m^2_{\eta}/M^2)$$

$$= \frac{8}{3} m_s^2\left[\frac{3}{8\pi^2} M^4\left[1 - \frac{\alpha_s}{\pi}\left\{\frac{5}{3} - \frac{22}{3}\left(\gamma - \ln\frac{M^2}{\mu^2}\right)\right\}\right] E_1\left(\frac{W^2}{M^2}\right) - 2m_s\left\langle \bar{s}s\right\rangle + \frac{1}{4}\left\langle\frac{\alpha_s}{\pi}G^2\right\rangle\right.$$

$$\left. + \frac{1}{M^2}\left\{\frac{16}{3}\pi^2\frac{\alpha_s}{\pi}\left\langle \bar{s}s\right\rangle^2 + m_s\left\langle \bar{s}g_s\sigma.Gs\right\rangle\right\}\right]. \qquad \ldots(40)$$

Notice that there is no division by $q^2$ as in the case of $P_L^{88}(q^2)$ and therefore there is no spurious pole. For the same reason $S(q^2)$ grows faster at large $q^2$ than $P_L^{88}(q^2)$ which means excited states are more significant in $S(q^2)$ than in $P_L^{88}(q^2)$. Also the residues at η and η' have additional $m_\eta^2$ and $m_{\eta'}^2$ respectively as compared to Eq. (38) . We use $W^2 = 2.3$ GeV$^2$ and the values of the parameters same as in Eq.(33). In order to reduce the contributions of excited states to a reasonable limit ( $\lesssim$ 32%), the limits on the Borel parameter was taken somewhat lower in this case : 0.6 GeV$^2 \leq M^2 \leq 0.9$ GeV$^2$. Our results of fits are displayed in Fig.2. We have

$$3\, m_\eta^4 \left(f_\eta^8\right)^2/(8\, m_s^2) = 3.64\times10^{-2}\ \text{GeV}^4,\quad 3\, m_{\eta'}^4 \left(f_{\eta'}^8\right)^2/(8\, m_s^2) = 4.02\times10^{-2}\ \text{GeV}^4\,. \qquad \ldots(41)$$

This corresponds to

$$f_8 = 168.4 \text{ MeV and } |\theta_8| = 18.9^\circ, \qquad \ldots(41a)$$

very close to the values listed in Eq. (38a). This confirms that our introduction of the spurious pole in Eq (37) is correct. In order to compare the quality of fits obtained from various curves, we define $\chi^2$ by the relation

$$\chi^2 = \left(\sum_{i=0}^{n}[f(x_i) - f_{fit}(x_i)]^2/[f(x_i) + f_{fit}(x_i)]^2\right)/(1+n).. \qquad \ldots(42)$$

The values corresponding to Fig.1 and Fig.2 are given in Table II. It is seen that as expected $P_L^{88}(q^2)$ fits better than $S(q^2)$. We can check the effect of changing the lower and higher $M^2$ ends on $\chi^2$. For Eq.(40), the interval 0.5 GeV$^2 \leq M^2 \leq$ 0.9 GeV$^2$ gives $f_8$= 165.5 MeV and $|\theta_8|$= 20.2° with $\chi$=2.5×10$^{-3}$, while 0.6 GeV$^2 \leq M^2 \leq$ 1.0 GeV$^2$ gives $f_8$= 171.7 MeV and $|\theta_8|$= 17.2° with $\chi$=3.1×10$^{-3}$ and excited states contribution rising to the level of 42% for the last case. For further discussion, we consider only the values given in Eq.(41a).

We next consider the sum rule for $P_L^{08}(q^2)$. We have

$$K^{08} + m_{\eta'}^2\, f_{\eta'}^0\, f_{\eta'}^8 \exp(-m_{\eta'}^2/M^2) + m_\eta^2\, f_\eta^0\, f_\eta^8 \exp(-m_\eta^2/M^2)$$

$$= \frac{3}{\sqrt{2}\pi^2}\left(\frac{\alpha_s}{\pi}\right)^2 m_s^2\, M^2 \left(\frac{7}{4} - \frac{1}{2}\ln\frac{M^2}{\mu^2}\right) E_0\left(W^2/M^2\right)$$

$$- \frac{1}{\sqrt{2}\pi^2}\, m_s^2\, M^2 \left\{1 + \frac{\alpha_s}{\pi}\left(\frac{17}{3} + 2\gamma - 2\ln\frac{M^2}{\mu^2}\right)\right\} E_0\left(W^2/M^2\right)$$

$$+ \frac{4\sqrt{2}}{3}\, m_s \left\langle \bar{s}s\right\rangle - 2\sqrt{2}\left(\frac{\alpha_s}{\pi}\right)^2 m_s \left\langle \bar{s}s\right\rangle \left(\gamma - \ln\frac{M^2}{\mu^2}\right) + \frac{\sqrt{2}}{3}\, m_s^2\, \frac{1}{M^2}\left\langle \frac{\alpha_s}{\pi}G^2\right\rangle$$

$$+ \frac{1}{\sqrt{2}}\frac{\alpha_s}{\pi}\, m_s^2\, \frac{1}{M^2}\left\langle \frac{\alpha_s}{\pi}G^2\right\rangle \left(1 - \gamma + \ln\frac{M^2}{\mu^2}\right) - \frac{8\sqrt{2}}{3}\, m_s^3\, \frac{1}{M^2}\left\langle \bar{s}s\right\rangle$$

$$- \sqrt{2}\,\frac{\alpha_s}{\pi}\, m_s\, \frac{1}{M^2}\left\langle \bar{s}g_s\sigma.Gs\right\rangle + \frac{32\sqrt{2}}{9}\pi^2\, \frac{\alpha_s}{\pi}\, m_s^2\, \frac{1}{M^4}\left\langle \bar{s}s\right\rangle^2$$

$$+\frac{2\sqrt{2}}{3}\, m_s^3\, \frac{1}{M^4}\left\langle \bar{s} g_s \sigma.G s\right\rangle . \qquad ...(43)$$

We again take $W^2$ = 2.3 $GeV^2$ with parameter values same as in Eq.(33). The details of the fit in the interval 0.8 $GeV^2 \leq M^2 \leq$ 1.5 $GeV^2$, are displayed in Fig.3 with the result

$K^{08} = -3.7\times 10^{-3}\; GeV^4$, $m_\eta^2 f_\eta^0 f_\eta^8 = 1.36\times 10^{-3}\; GeV^4$ and $m_{\eta'}^2 f_{\eta'}^0 f_{\eta'}^8 = -7.97\times 10^{-3}\; GeV^4$. …(44)

Since we have $f_\eta^0 = -f_0 \sin\theta_0$ positive and $f_{\eta'}^8 = f_8 \sin\theta_8$ negative, it follows that both $\theta_0$ and $\theta_8$ are negative. Combining with Eq.(38a) we find

$f_0$ = 142.3 MeV and $\theta_0 = -11.1^0$. .... (44a)

Let us now consider the combination $F(q^2) = -P_L^{00}(q^2) - 12\, \chi(q^2)/q^2$ where $\chi(q^2)$ is defined by Eq.(8). This has the effect of removing $\chi(q^2)$ from $-P_L^{00}(q^2)$, and has the advantage that the $F(q^2)$ receives no contribution for direct instantons. We can write the sum rule corresponding to $F(q^2)$ as

$$K - \frac{1}{2} m_{\eta'}^2 [(f_8 \sin\theta_8)^2 + 2\sqrt{2}\, f_0 f_8 \cos\theta_0 \sin\theta_8] e^{-m_{\eta'}^2/M^2}$$

$$-\frac{1}{2} m_\eta^2 [(f_8 \cos\theta_8)^2 - 2\sqrt{2}\, f_0 f_8 \sin\theta_0 \cos\theta_8] e^{-m_\eta^2/M^2}$$

$$= -\frac{3}{\pi^2}\left(\frac{\alpha_s}{\pi}\right)^2 m_s^2\, M^2 \left(\frac{7}{4} - \frac{1}{2}\ln\frac{M^2}{\mu^2}\right) E_0\left(W^2/M^2\right)$$

$$+\frac{1}{2\pi^2}\, m_s^2\, M^2 \left\{1 + \frac{\alpha_s}{\pi}\left(\frac{17}{3} + 2\gamma - 2\ln\frac{M^2}{\mu^2}\right)\right\} E_0\left(W^2/M^2\right)$$

$$-\frac{4}{3}\, m_s \left\langle \bar{s}s\right\rangle + 4\left(\frac{\alpha_s}{\pi}\right)^2 m_s \left\langle \bar{s}s\right\rangle \left(\gamma - \ln\frac{M^2}{\mu^2}\right) - \frac{1}{3}\, m_s^2\, \frac{1}{M^2}\left\langle \frac{\alpha_s}{\pi} G^2\right\rangle$$

$$-\frac{\alpha_s}{\pi}\, m_s^2\, \frac{1}{M^2}\left\langle \frac{\alpha_s}{\pi} G^2\right\rangle \left(1 - \gamma + \ln\frac{M^2}{\mu^2}\right) + \frac{8}{3}\, m_s^3\, \frac{1}{M^2}\left\langle \bar{s}s\right\rangle$$

$$+2\,\frac{\alpha_s}{\pi}\, m_s\, \frac{1}{M^2}\left\langle \bar{s} g_s \sigma.G s\right\rangle - \frac{32}{9}\,\pi^2\,\frac{\alpha_s}{\pi}\, m_s^2\, \frac{1}{M^4}\left\langle \bar{s}s\right\rangle^2$$

$$-\frac{2}{3}\, m_s^3\, \frac{1}{M^4}\left\langle \bar{s} g_s \sigma.G s\right\rangle . \qquad …(45)$$

Using $W^2$=2.3 $GeV^2$ and the same values of other parameters as in Eq.(33), we have fitted Eq.(45) in the range 0.8 $GeV^2 \leq M^2 \leq$ 1.5 $GeV^2$ and displayed it in Fig.4. This gives

K = $4.64\times 10^{-3}$ $GeV^4$, $-\frac{1}{2} m_\eta^2 [(f_8 \cos\theta_8)^2 - 2\sqrt{2}\, f_0 f_8 \sin\theta_0 \cos\theta_8] = -6.58\times 10^{-3}$ $GeV^4$, and

$-\frac{1}{2} m_{\eta'}^2 [(f_8 \sin\theta_8)^2 + 2\sqrt{2}\, f_0 f_8 \cos\theta_0 \sin\theta_8] = 9.2\times 10^{-3}$ $GeV^4$. …(46)

Combining with Eq.(38a) we find

$f_0$ = 140.5 MeV and $\theta_0 = -14.6^0$. … (46a)

We reconsider the sum rule for $\chi'(q^2)/q^2 - \chi'(0)/q^2$ from a slightly different perspective than in our earlier work[17] where we determined χ'(0) using the empirical values of the $f_8$, $f_0$, $\theta_8$, $\theta_0$ for residues of poles at η and η'. Here, we shall regard χ'(0) as well as the pole residues as unknowns to be determined by the sum rule. Writing it in the form (we set $m_u = m_d = 0$ so that the pion pole is absent)

$$\chi'(0) - \frac{1}{24}(f_8\cos\theta_8 - \sqrt{2}f_0\sin\theta_0)^2(1+\frac{m_\eta^2}{M^2})e^{-\frac{m_\eta^2}{M^2}} - \frac{1}{24}(f_8\sin\theta_8 + \sqrt{2}f_0\cos\theta_0)^2(1+\frac{m_{\eta'}^2}{M^2})e^{-\frac{m_{\eta'}^2}{M^2}}$$

$$= -(\frac{\alpha_s}{4\pi})^2\frac{1}{\pi^2}M^2E_0(\frac{w^2}{M^2})[1+\frac{\alpha_s}{\pi}\frac{74}{4}+\frac{\alpha_s}{\pi}\frac{9}{2}(\gamma - \ln\frac{M^2}{\mu^2})] - 16(\frac{\alpha_s}{4\pi})^3\frac{1}{M^2}m_s\left\langle \bar{s}s\right\rangle$$
$$-\frac{9}{64}\frac{1}{M^2}(\frac{\alpha_s}{\pi})^2\left\langle\frac{\alpha_s}{\pi}G^2\right\rangle + \frac{1}{16}\frac{1}{M^4}\frac{\alpha_s}{\pi}\left\langle g_s\frac{\alpha_s}{\pi}G^3\right\rangle - \frac{5}{128}\frac{\pi^2}{M^6}\frac{\alpha_s}{\pi}\left\langle\frac{\alpha_s}{\pi}G^2\right\rangle^2 .$$

…(47)

We have ignored the possible contribution from direct instantons given in the last term in Eq. (28). In Ref. [17] we had already pointed out that adding the direct instanton (DI) term without screening [21] gives an absurdly large contribution in Eq. (47) and completely destroys the sum rule. This point will be discussed below later , but for now, discard the plausible DI terms in Eq.(47) . With $W^2$ = 2.3 $GeV^2$ and other parameters same as in Eq.(33), fitting Eq. (47) in the range 0.8 $GeV^2 \le M^2 \le$ 1.5 $GeV^2$, we have from Fig.5

χ'(0) = $1.65\times10^{-3}$ $GeV^2$, …(48a)

$\frac{1}{24}(f_8\cos\theta_8 - \sqrt{2}f_0\sin\theta_0)^2 = 1.47\times10^{-3}$ $GeV^4$, …(48b)

$\frac{1}{24}(f_8\sin\theta_8 + \sqrt{2}f_0\cos\theta_0)^2 = 9.67\times10^{-4}$ $GeV^4$. …(48c)

On combining with the results from Eq.(38a) we get

$f_0$ = 152.5 MeV and $\theta_0 = -5.9^0$. …(48d)

Not surprisingly, when the values of $f_8$ , $f_0$ , $\theta_8$ and $\theta_0$ used in Ref. [17] are used in Eqs. (48b) and (48c) , the numbers obtained here are recovered. However, there is a small difference in the value of χ'(0), which can be accounted for by including pion pole contribution which was done in Ref.[17] but is ignored here.

Before we go on with remaining sum rules , we list in Table I , the values for $f_8$ , $f_0$ , $\theta_8$ and $\theta_0$ from the results of Eqs.(38a, 41a, 44a, 46a, 48d). We also note that from Table II, with values used in Eq.(33), the quality of fit is best for $P_L^{88}$ followed by $P_L^{08}$ , S, F and χ'.

For completeness we also consider the sum rule for $P_L^{00}(q^2)$ . We have

$$K^{00}+m_{\eta'}^2\left(f_{\eta'}^0\right)^2\exp(-m_{\eta'}^2/M^2)+m_{\eta}^2\left(f_{\eta}^0\right)^2\exp(-m_{\eta}^2/M^2)$$

$$=\frac{3}{8\pi^2}\left(\frac{\alpha_s}{\pi}\right)^2 M^4E_1\left(\frac{w^2}{M^2}\right)\left[1+\frac{\alpha_s}{\pi}\left\{\frac{65}{4}+\frac{9}{2}\left(\gamma-\ln\frac{M^2}{\mu^2}\right)\right\}\right]-\frac{3}{\pi^2}\left(\frac{\alpha_s}{\pi}\right)^2 m_s^2M^2\times$$

$$\left(\frac{7}{4}-\frac{1}{2}\ln\frac{M^2}{\mu^2}\right)E_0\left(\frac{W^2}{M^2}\right)+\frac{1}{2\pi^2}m_s^2M^2\left[1+\frac{\alpha_s}{\pi}\left(\frac{17}{3}+2\gamma-2\ln\frac{M^2}{\mu^2}\right)\right]E_0\left(\frac{W^2}{M^2}\right)$$

## TABLE I

Determination of the coupling constants of η and η′ mesons for the octet and singlet axial vector current with one or a combination of two equations out of Eqs. (38a), (41a), (44a), (46a), (48d).

| Eqs. used | Sum rule/ sum rule pair | $f_8\cos\theta_8$ (MeV) | $-f_8\sin\theta_8$ (MeV) | $f_0\cos\theta_0$ (MeV) | $-f_0\sin\theta_0$ (MeV) | $f_8$ (MeV) | $f_0$ (MeV) | $-\theta_8$ (degree) | $-\theta_0$ (degree) |
|---|---|---|---|---|---|---|---|---|---|
| 38a | $P_L^{88}$ | 165.6 | 62.2 | --- | --- | 176.8 | --- | 20.6 | --- |
| 41a | S | 159.3 | 54.6 | --- | --- | 168.4 | --- | 18.9 | --- |
| 38a,48d | $P_L^{88}$ , χ' | 165.6 | 62.2 | 151.7 | 15.7 | 176.8 | 152.5 | 20.6 | 05.9 |
| 38a,44a | $P_L^{88}$ $P_L^{08}$ | 165.6 | 62.2 | 139.6 | 27.5 | 176.8 | 142.3 | 20.6 | 11.1 |
| 38a,46a | $P_L^{88}$ , F | 165.6 | 62.2 | 136.0 | 35.4 | 176.8 | 140.5 | 20.6 | 14.6 |

## TABLE II

List of $\chi=\sqrt{\chi^2}$ of curves which are independent of instanton contribution. $\chi^2$ has been defined in Eq.(42). F has been defined in the text below Eq. (44a).

| Fig. no. | Plot of | χ | n |
|---|---|---|---|
| 1 | $P_L^{88}$ | $2.0\times10^{-4}$ | 28 |
| 2 | S | $1.6\times10^{-3}$ | 28 |
| 3 | $P_L^{08}$ | $1.1\times10^{-3}$ | 28 |
| 5 | χ' | $5.1\times10^{-3}$ | 28 |
| 4 | F | $2.8\times10^{-3}$ | 28 |

$$-\frac{4}{3}m_s\langle\bar{s}s\rangle+\frac{3}{4}\frac{\alpha_s}{\pi}\left\langle\frac{\alpha_s}{\pi}G^2\right\rangle\left[1-\frac{9}{4}\frac{\alpha_s}{\pi}\left(\ln\frac{M^2}{\mu^2}-\gamma\right)\right]+4\left(\frac{\alpha_s}{\pi}\right)^2 m_s\langle\bar{s}s\rangle\left(\gamma-\ln\frac{M^2}{\mu^2}\right)$$

$$-\frac{\alpha_s}{\pi}\frac{m_s^2}{M^2}\left\langle\frac{\alpha_s}{\pi}G^2\right\rangle\left(1-\gamma+\ln\frac{M^2}{\mu^2}\right)-\frac{1}{3}\frac{m_s^2}{M^2}\left\langle\frac{\alpha_s}{\pi}G^2\right\rangle+\frac{3}{2}\frac{\alpha_s}{\pi}\frac{1}{M^2}\left\langle g_s\frac{\alpha_s}{\pi}G^3\right\rangle$$

$$+2\frac{\alpha_s}{\pi}\frac{m_s}{M^2}\langle \bar{s}g_s\sigma.Gs\rangle + \frac{8}{3}\frac{m_s^3}{M^2}\langle \bar{s}s\rangle - 3\left(\frac{\alpha_s}{\pi}\right)^3 m_s\langle \bar{s}s\rangle\left(\frac{1}{2}-\gamma+\ln\frac{M^2}{\mu^2}\right)$$

$$+\frac{45}{64}\frac{\pi^2}{M^4}\frac{\alpha_s}{\pi}\left\langle\frac{\alpha_s}{\pi}G^2\right\rangle^2 - \frac{32}{9}\pi^2\frac{\alpha_s}{\pi}m_s^2\frac{1}{M^4}\langle \bar{s}s\rangle^2 - \frac{2}{3}\frac{m_s^3}{M^4}\langle \bar{s}g_s\sigma.Gs\rangle$$

+ direct instanton + screening terms. …(49)

A brief discussion of the last two terms of Eq.(49) is now necessary . Although there is no universally accepted description of the QCD vacuum, the model based on instanton fluid, which regards the ground state a collection of instanton –anti-instanton pairs has been widely used to study a number of vacuum correlation functions [23]. As is well-known, an instanton of size ρ located at $x_0$ [ see Eq.(65) of Forkel [21] ] corresponds to the field strength

$$G_{\mu\nu}^{(I),a}(x) = \frac{-4\rho^2}{g_s}\frac{\eta_{a\mu\nu}}{\left[(x-x_0)^2+\rho^2\right]^2}$$

where $\eta_{a\mu\nu}$ is a 't Hooft symbol . This means the anomaly- anomaly correlation has a contribution directly from the distribution of intstantons in the vacuum. In the picture in which instantons are non-interacting, the calculation is simple as in statistical mechanics of a non-interacting gas. In Eq. (29) the DI contribution, for the constant density case is displayed, and in Eq.(30), the density function in the Gaussian tail approximation [21] for using in the numerical integration in Eq.(28) to get the DI contribution is displayed. Forkel has pointed out that this contribution has to be corrected for screening caused by exchanges of the Goldstone fields. We then begin first ignoring both the direct instanton term and screening in the right hand side of Eq.(49); this and the contribution of DI with density in Gaussian tail-approximation of Eq (30a) are shown in Fig.6. It is easily seen that DI term is much too large. Forkel has estimated the screening corrections arising from η , η′ exchange as

$$i\int d^4x e^{iqx}\langle 0|T\{Q_5(x), Q_5(0)\}|0\rangle_{DISC} = (8\pi)^2\left(\frac{F_{\eta'}^2}{Q^2+m_{\eta'}^2}+\frac{F_\eta^2}{Q^2+m_\eta^2}\right), \quad (Q^2=-q^2) \qquad \dots(50)$$

where the subscript in the LHS refers to the screening and

$F_\eta^2$ = 0.0876 GeV$^6$, $F_{\eta'}^2$ = 0.543 GeV$^6$. …(51)

In Fig.7, the DI term and the screening term are displayed after Borel-transformation to $M^2$ which shows that the screening is comparable to the DI . We have already seen that the sum rule for $\chi'(q^2)/q^2$, Eqs. (47) and (48d), works very well by discarding the DI and screening and more importantly yields values for the couplings consistent with values obtained from Eqs (38a) and (44a) which have no direct instanton terms at all. Encouraged by this, we can consider the possibility of the screening term in $P_L^{00}(q^2)$ being even larger than DI . To be specific we tried the form

DI + Screening = δ×RHS[Eq.(30a)] ….(52)

where δ is some numerical factor to be determined by fitting Eq.(49). We find the value δ = −0.074 fits the sum rule well as can be seen for Figs. 8 ( curves B and C ).

Taking $W^2$=2.5 GeV$^2$ and the values of the parameters as in Eq.(33), we fit the sum rule in the range 0.8GeV$^2$<$M^2$<1.5GeV$^2$ with results

$K^{00} = -9.22\times10^{-4}$ GeV$^4$, … (53a)

$m_\eta^2 \left(f_\eta^0\right)^2 = 7.2\times10^{-5}$ GeV$^4$, … (53b )

$m_{\eta'}^2 \left(f_{\eta'}^0\right)^2 = 1.812\times10^{-2}$ GeV$^4$. …(53c)

This corresponds to $f_0$ = 141.4 MeV and $\theta_0 = \pm6.4^0$. To see the sensitivity of the physical parameters on the coefficients of the DI term taken, we find that a change of δ from −0.074 to −0.1, changes the results substantially

$K^{00} = -6.872\times10^{-3}$ GeV$^4$, $m_\eta^2 \left(f_\eta^0\right)^2 = 6.433\times10^{-3}$ GeV$^4$, $m_{\eta'}^2 \left(f_{\eta'}^0\right)^2 = 1.866\times10^{-2}$ GeV$^4$,

which corresponds to $f_0$ = 204.5 MeV and $\theta_0 = \pm45.8^0$.

We now turn to the sum rule for $\chi(q^2)/q^2 - \chi(0)/q^2$. We have

$$-\chi(0) + \frac{m_\eta^2}{24}(f_8\cos\theta_8 - \sqrt{2}f_0\sin\theta_0)^2 e^{-\frac{m_\eta^2}{M^2}} + \frac{m_{\eta'}^2}{24}(f_8\sin\theta_8 + \sqrt{2}f_0\cos\theta_0)^2 e^{-\frac{m_{\eta'}^2}{M^2}}$$

$$= (\frac{\alpha_s}{4\pi})^2 \frac{1}{2\pi^2} M^4 E_1(\frac{w^2}{M^2})[1 + \frac{\alpha_s}{\pi}(\frac{65}{4} + \frac{9}{2}(\gamma - \ln\frac{M^2}{\mu^2})]$$

$$+ \frac{1}{16}\frac{\alpha_s}{\pi}\left\langle \frac{\alpha_s}{\pi}G^2 \right\rangle \left[1 - \frac{9}{4}\frac{\alpha_s}{\pi}\left(\ln\frac{M^2}{\mu^2} - \gamma\right)\right] + \frac{1}{8}\frac{1}{M^2}\frac{\alpha_s}{\pi}\left\langle g_s \frac{\alpha_s}{\pi}G^3 \right\rangle$$

$$+ \frac{15}{256}\frac{\pi^2}{M^4}\frac{\alpha_s}{\pi}\left\langle \frac{\alpha_s}{\pi}G^2 \right\rangle^2 - 16\left(\frac{\alpha_s}{4\pi}\right)^3 m_s \left\langle \bar{s}s \right\rangle \left(\frac{1}{2} + \ln\frac{M^2}{\mu^2} - \gamma\right)$$

$$-0.074 \times \hat{B}\left[-\frac{1}{2}q^2 \int d\rho n(\rho)\rho^4 K_2^2(Q\rho)\right] \quad \ldots(54)$$

For $P_L^{00}(q^2)$, Eq.(49) above, we accounted for screening by multiplying the DI by δ = −0.074. As explained already with other sum rules, replacing T-product by OPE means χ(0) may not satisfy Eq( A.2) as demanded by low energy theorems. As in the analysis of sum rule (49), we again take $W^2 = 2.5$ GeV$^2$ and fit in the range 0.8 GeV$^2 \leq M^2 \leq 1.5$GeV$^2$. We get, cf. Fig(9) and Fig.(10):

$\chi(0) = 4.3\times10^{-4}$ GeV$^4$, …(55a)

$m_\eta^2 \frac{1}{24}(f_8\cos\theta_8 - \sqrt{2}f_0\sin\theta_0)^2 = 4.4\times10^{-4}$ GeV$^4$, …(55b)

$m_{\eta'}^2 \frac{1}{24}(f_8\sin\theta_8 + \sqrt{2}f_0\cos\theta_0)^2 = 8.5\times10^{-4}$ GeV$^4$. … (55c)

as the constant and residues at the η- and η′-poles respectively. This corresponds to $f_0$ =150.2 MeV and $\theta_0 = -6.0^0$ assuming the values of $f_8$ and $\theta_8$ as given by $P_L^{88}(q^2)$ - sum rule.

It is instructive to compare the results for the $\chi'(q^2)/q^2$, $P_L^{00}(q^2)$ and $\chi(q^2)/q^2$ as given in Figs (5, 8, 10 ). First, although the addition of small DI with a negative coefficient −0.074 is ad hoc, the same factor fits both Eq.(54) and Eq.(49) reasonably. Moreover, η and η′ residues are in the ratio 4.4: 8.5 in Eqs. (55b) and (55c) , that is, roughly a factor of two. In the $P_L^{00}(q^2)$ sum rule, the η residue is very small compared to η′ residue as seen in Eqs. (53b) and (53c), with values $7.2\times10^{-5}$ GeV$^4$ and $1.812\times10^{-2}$ GeV$^4$ respectively, that is, differing roughly by a factor of 250. Let us now compare the residue results of $\chi'(q^2)$ and $\chi(q^2)$ . In the former we have discarded the DI and screening assuming them to cancel each other, while in the latter, the screening is slightly larger as reflected by the factor −0.074 . We have from Eqs. (48b) and (48c)

$$\frac{1}{24}(f_8\cos\theta_8-\sqrt{2}f_0\sin\theta_0)^2 = 1.47\times10^{-3}\ \text{GeV}^4, \qquad \ldots(48b)$$

$$\frac{1}{24}(f_8\sin\theta_8+\sqrt{2}f_0\cos\theta_0)^2 = 9.67\times10^{-4}\ \text{GeV}^4, \qquad \ldots(48c)$$

while from Eqs. (55b) and (55c)

$$\frac{1}{24}(f_8\cos\theta_8-\sqrt{2}f_0\sin\theta_0)^2 = 1.47\times10^{-3}\ \text{GeV}^4,$$

$$\frac{1}{24}(f_8\sin\theta_8+\sqrt{2}f_0\cos\theta_0)^2 = 9.26\times10^{-4}\ \text{GeV}^4.$$

which are remarkably close. It therefore appears to conclude a) Screening corrections to DI are vital to obtain consistent results from sum rules . b) For the derivative of the topological susceptibility, the screening is almost complete, at least given the uncertainties inherent in the sum rule approach. It is useful to compare the coefficients $F_\eta^2$ = 0.0886 GeV$^6$ and $F_{\eta'}^2$ = 0.543 GeV$^6$ used by Forkel[21] with the matrix element of the anomaly between the vacuum and pseudoscalar states. From the equations [24]

$$\langle 0|\frac{3\alpha_s}{4\pi}G\tilde{G}|\eta\rangle = \sqrt{\frac{3}{2}}m_\eta^2(f_8\cos\theta_8-\sqrt{2}f_0\sin\theta_0),$$

$$\langle 0|\frac{3\alpha_s}{4\pi}G\tilde{G}|\eta'\rangle = \sqrt{\frac{3}{2}}m_{\eta'}^2(f_8\sin\theta_8+\sqrt{2}f_0\cos\theta_0),$$

we have from Eq. (48b) and Eq.(48c)

$$\langle 0|\alpha_s G\tilde{G}|\eta\rangle\langle\eta|\alpha_s G\tilde{G}|0\rangle = 0.083GeV^6$$

and

$$\langle 0|\alpha_s G\tilde{G}|\eta'\rangle\langle\eta'|\alpha_s G\tilde{G}|0\rangle = 0.515GeV^6$$

which are close to the numbers of Forkel [21] used by us in our Eqs.(50) and (51) above. We must also add that since u and d quark masses are different, $\langle 0|\alpha_s G\tilde{G}|\pi^0\rangle \neq 0$ and pion exchange contributes to screening. In Fig.7 and Fig.11, we have displayed the specific values of Forkel [21]

who ignores the pion. While accepting the general picture, we can not be quantitatively accurate. We emphasize that screening effects require more study.

## V. SUMMARY AND CONCLUDING REMARKS

We have considered seven functions consisting of axial current correlators and pseudoscalar current correlators: $P_L^{88}(q^2)$, S($q^2$), $P_L^{08}(q^2)$, $-P_L^{00}(q^2)-12\frac{\chi(q^2)}{q^2}$, $\frac{\chi'(q^2)}{q^2}$, $P_L^{00}(q^2)$ and $\frac{\chi(q^2)}{q^2}$ and corresponding sum rules. The first four have no contributions from direct instantons , while the last three would have possible contributions. The octet current couplings are well determined by the first two functions, and as expected , $P_L^{88}(q^2)$ sum rule works better with a better $\chi^2$ than S($q^2$). As displayed in Table I , both sum rules give nearly same values for the octet coupling and the mixing angle. With the knowledge of the octet couplings, we have seen that the octet-singlet correlator $P_L^{08}(q^2)$, works better than the hybrid function F. While the feature that the sign of both angles is negative and that singlet coupling and the magnitude of the singlet angle are smaller than the octet counterparts is true, $P_L^{08}(q^2)$ sum rule results are closer to phenomenological values than the F sum rule results.

As noted in our earlier work [17], we have that sum rule for $\frac{\chi'(q^2)}{q^2}$ without any direct instantons works very well. We used this observation and a semi-quantitative discussion of screening of direct instanton, to find a simple multiplicative factor to get reasonable fits of $P_L^{00}(q^2)$ and $\frac{\chi(q^2)}{q^2}$ sum rules. We have pointed out that while division by $q^2$ improves asymptotic behavior and therefore gives better sum rules, it can introduce a spurious pole at $q^2$ =0 and should be accounted for in the analysis. We found that constants $K^{88}$, $K^{08}$, etc. are not zero as demanded by low energy theorems. This caveat also applies to $\chi'(0)$. However, as discussed in the introduction, the $\frac{\chi'(q^2)}{q^2}$ sum rule value is close to values of three other determinations, namely axial current sum rules [18], Bjorken sum rule [18, 14] and chiral perturbation theory [19].

As pointed out earlier, sum rule determinations are subject to errors arising from the uncertainties in VEV's of various operators, the values of $W^2$ - the continuum threshold, variation in match region of the Borel mass variable, in the ignored higher dimensional terms in the OPE and higher order terms in the Wilson coefficients. It is usual to expect that the errors are in the (10-15)% range. Nevertheless, we can rely on our results since these are mutually consistent and are also in agreement with phenomenological values as seen from Table III. We emphasize that we stayed with the rules for the Borel mass range, which is limited at the lower end by the contribution of the highest dimensional terms on OPE, and at the higher end by the contributions of the excited states which we have limited by about 32% or less. We have uniformly used a value of 2.3 GeV$^2$ for W$^2$, except in the sum rules for $\chi$ and $P_L^{00}$, where we have used W$^2$=2.5 GeV$^2$, a slightly higher value to get a better fit. There are suggestions [25,26] that the increase in W$^2$ is necessitated for richer crops of resonances in the singlet channel as compared to the octet. Alternatively, violations of duality for singlet continuum states may be more important compared to the octet.

We summarize our results as follows. As noted in Table I , the values of $f_8$ and $\theta_8$ obtained from Eqs. (38a) and (41a) listed in the first two rows are close and certainly within the errors of the sum rule method. In Table I , we have given in the fifth and sixth columns the values for $f_0 \cos\theta_0 = f_{\eta'}^0$ and $-f_0 \sin\theta_0 = f_\eta^0$ obtained from using the results of the equations listed in the first column . We note that the value of $f_0 \cos\theta_0$ is better determined than $f_0 \sin\theta_0$. Part of the reason , is due to the different functional relations of the couplings at the η and η′ poles as seen from Eqs. (48b), (48c), (44) and (46). Despite this, the general feature that $f_0$ is smaller that $f_8$ and the numerical values of $\theta_0$ is significantly smaller than $\theta_8$ clearly emerges. In Table III we have listed the values of $f_8$, $\theta_8$, $f_0$ and $\theta_0$ from our work, the simple average of Eqs. (38a) and (41a) for $f_8$ and $\theta_8$ namely 172.6 MeV and $-19.8^{\circ}$ and these numbers are, in turn, used in Eqs. (48b), (48c) , (44) and (46) to obtain the average values for $f_0$ = 149.1 MeV and $\theta_0 = -10.9^{\circ}$.

In Table III, we have listed some of the results obtained in the current literature. Feldman and Kroll [27,28], using two-angle parameterization, have achieved a simultaneous description of the two-photon decays of η and η′ and the transition form factors of ηγ and η′γ at large momentum transfer. Shore[10] has derived the QCD formula for the two-photon decays of η and η′ and the corresponding DGMOR relations by generalizing conventional PCAC to include the effect of the anomaly in a way which is consistent with the renormalization group and 1/N$_C$ expansion. In Refs. [8,9]], the reader will find 1/N$_C$ expansion results in the context of chiral perturbation theory. It will be interesting to study the comparison of sum rule results with chiral perturbation theory. To conclude the discussion of Table III , we comment briefly on the results of References [16 , 15 and 10] which are listed on the last three rows of Table III . De Fazio and Pennington [16] had used a somewhat oversimplified approach to sum rules. To calculate η couplings they use the perturbative term without radiative corrections and only the $\langle \bar{q}q \rangle$ for the nonperturbative term with a low energy value for W$^2$. Moreover, to find η' coupling they simply increase the value of W$^2$. No details of combined fits to η and η' are given by them . Their results for the octet and singlet angles, which we have listed in Table III, are in disagreement with the earlier rows in Table III. We may add that Ref.[16] is also internally inconsistent as the value obtained for $\theta_8$ from pseudoscalar densities is $-23^0$ (not mentioned by them ) is different from $-8.4^0$ quoted and obtained by them using axial vector current correlators, unlike our results displayed in the first two rows of Table I. Turning to Ref. [15], we have already commented extensively in Ref. [17]. Briefly, the authors in Ref.[15] erroneously use physical η' mass instead of its value in the chiral limit, besides the fact that in their sum rule, the two sides hardly match with each other. Coming to the work of Shore in Ref.[10] , we note that it is based on generalized current algebra and is different from others listed in Table III. Ref.[10] has used the De Vechhia- Veneziano [31] formula [ their Eq.(A4′)]

$$\chi(q^2) = -\frac{aF_\pi^2}{2N_c}\left[1-\frac{a}{N_c}\sum_i \frac{1}{q^2-\mu_i^2}\right]^{-1} \qquad \ldots(56)$$

## TABLE III

Comparison of our results on couplings and mixing angles with those obtained by other authors. The values in the first row give the average of Eqs.(38a) and (41a) for $f_8$ and $\theta_8$. This, in turn, is used in Eqs (44) , (46) and Eqs.(48b), (48c) to obtain the average values of $f_0$ and $\theta_0$.

| Ref. | Specification | $f_8$ (MeV) | $f_0$ (MeV) | $\theta_8$ (Degree) | $\theta_0$ (Degree) |
|---|---|---|---|---|---|
| This work | Sum rules, averaged results | 172.6 | 149.1 | −19.8 | −10.9 |
| [29] | $f_8$ from [8]<br>$f_8$ from [7]<br>best fit phen. | 1.28 $f_\pi$=167.3<br>1.34 $f_\pi$=175.5<br>(1.51 ± 0.05) $f_\pi$ =197.4 ± 6.5 | 154.23 ± 5.2<br>156.84 ± 5.2<br>168.60 ± 5.2 | − (22.2 ± 1.8)<br>− (22.9 ± 1.8)<br>− (23.8 ± 1.4) | − (8.7 ± 2.1)<br>− (6.9 ± 2.0)<br>− (2.4 ± 1.9) |
| [27, 28] | Theory<br>Phen. | 155.53 ± 7.8<br>164.68 ± 7.8 | 143.77 ± 5.2<br>152.92 ± 5.2 | − (19.4 ± 1.4)<br>− (21.2 ± 1.4) | − (6.8 ± 1.4)<br>− (9.2 ± 1.4) |
| [8] | ChPT | 167.30 | 143.77 | −20.5 | −4.0 |
| [30] | ChPT | 172.53 | 164.05 | −20.0 | −1.0 ± 1.5 |
| [16] | Sum rules | 188.21 | 176.45 | −8.4 | −13.8 |
| [15] | Sum rules | | 178 ± 17 | − (17.0 ± 5.0) | |
| [10] | Extended current algebra | 148.0 | 150.7 | −20.1 | −12.3 |

where the Goldstone boson mass squared $\mu_i^2$ are related to quark condensate by

$$\mu_i^2 = -2m_i \frac{1}{F_\pi^2}\langle 0|\bar{q}q|0\rangle \qquad \ldots.(57)$$

and $a$ is some constant . Shore [10] uses Eq.(56) to get

$$\chi(0) = -A\left(1 - A\sum_{q=u,d,s}\frac{1}{m_q\langle\bar{q}q\rangle}\right)^{-1}$$

where the constant $A$ is

$$A = \frac{aF_\pi^2}{2N_c} \qquad \ldots(58)$$

and uses it in the DGMOR relation for the singlet sector to obtain

$$\left(f_\eta^0 m_\eta\right)^2 + \left(f_{\eta'}^0 m_{\eta'}\right)^2 = -\frac{2}{3}\left(m_u\left\langle \bar{u}u\right\rangle + m_d\left\langle \bar{d}d\right\rangle + m_s\left\langle \bar{s}s\right\rangle\right) + 6A \quad . \qquad \ldots(59)$$

Ioffe has correctly pointed out that Eq.(56) has a wrong pole structure at the Goldstone states. The reader can check by comparing Eq.(A4) in our Appendix with Eq.(56) written above that while $\chi(q^2)$ should have poles at the Goldstone states, Eq.(56) has zeroes at the Goldstone states and is therefore incorrect. Consequently the Eqs. (6.1), (6.4) and (6.5) used by Shore[10] can not be trusted.

## ACKNOWLEDGEMENT

J.P.S. thanks members of CHEP, IISc, Bangalore, for the hospitality at the Center where part of this work was done.

# APPENDIX

## Low energy theorems and regularity of functions used in sum rules at $q^2$=0

We first note that the anomaly-anomaly correlator or the topological susceptibility

$$\chi(q^2) = i\int d^4x e^{iqx}\left\langle 0\right| T\{Q_5(x), Q_5(0)\}\left|0\right\rangle \qquad \ldots(A1)$$

satisfies a low energy theorem. It was pointed out by Crewther [4] that $\chi(0)$ vanishes in any theory which has at least one massless quark. The large $N_c$ (number of colors) limit was considered by Veneziano[5]. They showed that in a theory with $N_f$ light quarks with masses $m_i << M$, where M is the mass of strong interaction

$$\chi(0) = -\left\langle 0\right|\bar{q}q\left|0\right\rangle \left(\sum_{i=1}^{N_f}\frac{1}{m_i}\right)^{-1} \qquad \ldots(A2)$$

Here $-\left\langle 0\right|\bar{q}q\left|0\right\rangle$ is the flavor symmetric value of the quark condensate and corrections of the order ($m_i$/M) have been neglected in Eq.(A2). Clearly the reduced mass

$$m_{red} = \left(\sum_{i=1}^{N_f}\frac{1}{m_i}\right)^{-1} \qquad \ldots(A3)$$

vanishes when any one of the $m_i$ is zero, consistent with Crewther's theorem [4]. Leutwyler and Smilga [32] were able to show that for the case of two light quarks Eq.(A2) is valid at any $N_c$. This was further extended for three flavors by Smilga[33].

The function $\chi(q^2)$ can be found for small $q^2$ using chiral perturbation theory[19]. It has the expansion

$$\chi(q^2) = \sum_{P=\pi^0,\eta}\frac{\left|\left\langle 0\right|Q_5\left|P\right\rangle\right|^2}{M_P^2 - q^2} - \frac{1}{9}BF^2(m_u+m_d+m_s) + \frac{1}{6}\tilde{H}_0 q^2 + O(q^4) \qquad \ldots(A4)$$

Here

$$\langle 0|Q_5|\pi^0\rangle = \frac{1-\frac{4}{3}\sin^2\varepsilon}{2\cos\varepsilon}(m_d - m_u)BF \qquad \text{...(A5)}$$

$$\langle 0|Q_5|\eta\rangle = \frac{2(1-4\sin^2\varepsilon)\cos\varepsilon}{3\sqrt{3}\cos 2\varepsilon}(m_s - \hat{m})\,\mathrm{BF} \qquad \text{...(A6)}$$

with

$$\hat{m} = \frac{1}{2}(\mathrm{m_u+m_d});\ \mathrm{BF}^2 = -\langle 0|\bar{q}q|0\rangle \qquad \text{....(A7)}$$

F is the flavor symmetric pseudoscalar decay constant (the pion decay constant $F_\pi$=92.4 MeV) and

$$M_{\pi^0}^2 = (\mathrm{m_u+m_d})\mathrm{B} - \Delta, \qquad \text{...(A8)}$$

$$M_\eta^2 = \frac{2}{3}(\hat{m}+2m_s)\,\mathrm{B} + \Delta \qquad \text{...(A9)}$$

$$\Delta = \frac{4\sin^2\varepsilon}{3\cos 2\varepsilon}(m_s - \hat{m})\,\mathrm{B}, \quad \tan 2\varepsilon = \frac{\sqrt{3}}{2}\frac{m_d - m_u}{m_s - \hat{m}}, \qquad \text{...(A10)}$$

$$\tilde{H}_0 = \mathrm{H_0 + F^2} \qquad \text{...(A11)}$$

$H_0$ is a constant appearing in the effective chiral Lagrangian. The effective action is obtained as a functional of external sources that couple to vector and axial vector currents, scalar and pseudoscalar densities and $Q_5(x)$ the winding number density. If θ(x) is the external source that couples to $Q_5(x)$ then H is the coefficient of the term quadratic in $\partial_\mu\theta(x)$ in an expansion of the effective action as a series in θ(x) and its derivatives.

It is important to note that $\chi(q^2)$ as given in Eq.(A4) contains the momentum independent term $-\frac{1}{9}\mathrm{B\,F^2(m_u+m_d+m_s)}$. Setting $q^2$=0 in Eq.(A4) one obtains

$$\chi(0) = -\,\mathrm{B\,F^2\,m_{red}} + \mathrm{O(m^2)} \qquad \text{...(A12)}$$

where $m_{red}$ is as given by Eq.(A3).

Ioffe [14] has derived the result (A12) above from yet another perspective; we briefly outline his derivation since it is useful in the context of understanding the regularity of the functions used for sum rules at $q^2$=0. Consider the singlet-singlet current correlator $\Pi_{\mu\nu}^{00}(q)$. Since there are no poles at $q^2$=0 in the physical correlator, we have

$$\lim_{q_\mu\to 0} q^\mu\,\Pi_{\mu\nu}^{00}(q)\,q^\nu = \lim_{q_\mu\to 0} -\mathrm{P}_L^{00}(q^2)\,q^2 \qquad \text{...(A13)}$$

$$= 0 \qquad \text{...(A14)}$$

which implies that $\mathrm{P}_L^{00}(q^2)$ is regular at $q^2$=0. On the other hand

$$\lim_{q_\mu\to 0} q^\mu\,\Pi_{\mu\nu}^{00}(q)\,q^\nu = i12\int d^4x e^{iqx}\langle 0|\mathrm{T}\{Q_5(x), Q_5(0)\}|0\rangle$$

$$+\, i2\int d^4x e^{iqx}\langle 0|\mathrm{T}\{Q_5(x), D(0)\}|0\rangle$$

$$+\, i2\int d^4x e^{iqx}\langle 0|\mathrm{T}\{D(x), Q_5(0)\}|0\rangle$$

$$+ i\,\frac{1}{3}\int d^4x e^{iqx}\left\langle 0\right| T\{D(x), D(0)\}\left|0\right\rangle + \frac{4}{3}\sum_{i=u,d,s} m_i\left\langle 0\right|\bar{q}_i q_i\left|0\right\rangle$$

$$= 0 \qquad \ldots(A15)$$

A plausible Schwinger term $\left[J^0_{05}(x), Q_5(0)\right]\delta(x_0)$ can be shown to be zero [14]. Similarly by considering the correlator

$$P_\mu(q) = i\int d^4x e^{iqx}\left\langle 0\right|\left[J^0_{\mu 5}(x), Q_5(0)\right]\left|0\right\rangle \qquad \ldots(A16)$$

and the fact that

$$\lim_{q_\mu \to 0} q^\mu P_\mu(q) = 0, \qquad \ldots(A17)$$

one derives

$$i\int d^4x\left\langle 0\right| T\{2Q_5(x), Q_5(0)\}\left|0\right\rangle + i\frac{1}{3}\int d^4x\left\langle 0\right| T\{D(x), Q_5(0)\}\left|0\right\rangle = 0 \qquad \ldots(A18)$$

Combining (A15) and (A18) one gets [14]

$$i12\int d^4x\left\langle 0\right| T\{Q_5(x), Q_5(0)\}\left|0\right\rangle - i\,\frac{1}{3}\int d^4x\left\langle 0\right| T\{D(x), D(0)\}\left|0\right\rangle$$

$$- \frac{4}{3}\sum_{i=u,d,s} m_i\left\langle 0\right|\bar{q}_i q_i\left|0\right\rangle = 0 \qquad \ldots(A19)$$

Ioffe rewrites Eq.(A19) in the form

$$i12\int d^4x\left\langle 0\right| T\{Q_5(x), Q_5(0)\}\left|0\right\rangle$$

$$= i\,\frac{1}{3}\int d^4x\left\langle 0\right| T\{D(x), D(0)\}\left|0\right\rangle + \frac{4}{3}\sum_{i=u,d,s} m_i\left\langle 0\right|\bar{q}_i q_i\left|0\right\rangle . \qquad \ldots(A20)$$

The term linear in the quark masses in the first term in the right hand side of Eq.(A20) can be found from the matrix elements of $\left\langle 0\right| D(x)\left|\pi^0\right\rangle$ and $\left\langle 0\right| D(x)\left|\eta\right\rangle$ and leads back to Eq.(A12). Complete details can be found in Ioffe [14]. For the purpose of the present paper, the important question is how the various terms in the right hand side of Eq.(A15) conspire to keep their sum to be zero, so that $P^{00}_L(q^2)$ is regular at $q^2 = 0$. We have seen above that the low energy theorem (A12), $\left|\pi^0\right\rangle$, $\left|\eta\right\rangle$ contributions and the equal time comutator add to give the zero. On the other hand, to derive the QCD sum rule for $P^{00}_L(q^2)$ we have operator product expansion for various terms like $i\int d^4x e^{iqx}\left\langle 0\right| T\{Q_5(x), Q_5(0)\}\left|0\right\rangle$ and $i\int d^4x e^{iqx}\left\langle 0\right| T\{Q_5(x), D(0)\}\left|0\right\rangle$ as given in Eq.(28) and Eq.(31) respectively, which is a good approximation at high $q^2$. We cannot, therefore, expect $q^2 P^{00}_L(q^2)$ = OPE + ETCR to vanish at $q^2 = 0$ as demanded by the low energy theorem. So in dividing by $q^2$ to derive an expression for $P^{00}_L(q^2)$ valid at large $q^2$, we introduce a spurious pole at $q^2 = 0$.

We can explicate this by considering the octet-octet correlator in some detail. To simplify matters we set $m_u = m_d = 0$ but keep $m_s \neq 0$. Isospin is exact in this limit, so that the current $J^{8}_{\mu 5}(x)$ does not couple to the pion and therefore the correlator $\pi^{88}_{\mu\nu}$ is still regular at $q^2 = 0$. Consider now the analog of Eq.(A15). We have

$$\lim_{q_\mu \to 0} q^{\mu}\, \Pi^{88}_{\mu\nu}(q)\, q^{\nu} = i\frac{4}{6} \int d^4x \langle 0|\,T\{D_s(x), D_s(0)\}\,|0\rangle + \frac{8}{3} m_s \langle 0|\bar{s}s|0\rangle \qquad \ldots(A21)$$

Here $D_s(x) = 2i\, m_s \bar{s}(x)\gamma_5 s(x)$. To see how the two terms in Eq.(A21) cancel upto linear order in $m_s$ we note

$$\langle 0|D_s|\eta\rangle = -\sqrt{\frac{3}{2}} f_\pi m_\eta^2 \qquad \ldots(A22)$$

so that

$$\frac{4}{6} i \int d^4x \langle 0|\,T\{D_s(x), D_s(0)\} = f_\pi^2 m_\eta^2 + \text{higher order terms.} \qquad \ldots(A23)$$

Now by GMOR relation

$$f_\pi^2 m_\eta^2 = -\frac{8}{3} m_s \langle 0|\bar{s}s|0\rangle . \qquad \ldots(A24)$$

so the right hand side of Eq.(A21) adds to zero, thus preserving the regularity of $P^{88}_L(q^2)$ at $q^2 = 0$. Returning to the QCD sum rule for $P^{88}_L(q^2)$, we have from Eq.(34)

$$-q^2 P^{88}_L(q^2) = \frac{4}{6} i \int d^4x e^{iqx} \langle 0|\,T\{D_s(x), D_s(0)\}\,|0\rangle + \frac{8}{3} m_s \langle 0|\bar{s}(0)s(0)|0\rangle \qquad \ldots(A25)$$

$$\underset{|q^2|\to\infty}{\approx} \left\{ -\frac{1}{\pi^2} m_s^2 q^2 \left[ \ln\left(\frac{-q^2}{\mu^2}\right) - 2 \right] + \ldots \right\} + \frac{8}{3} m_s \langle 0|\bar{s}s|0\rangle \qquad \ldots(A26)$$

where we have used the operator expansion for $T\{D_s(x), D_s(0)\}$. Unlike the r.h.s. of Eq.(A25) which vanishes as $q_\mu \to 0$, we cannot expect the r.h.s. of Eq.(A26) to vanish at $q^2 = 0$. Apart from the fact that in Eq.(A26) we are using large $q^2$ approximation in the actual sum rule evaluation we also use numerical estimates for $m_s$ and $\langle 0|\bar{s}s|0\rangle$ while in Eq.(A25) we used an algebraic identity using current algebra. Therefore in the process of dividing by $q^2$ in Eq.(A26) we introduce a spurious pole at $q^2 = 0$ in $P^{88}_L(q^2)$ which must be accounted for. On Borel transformation the spurious pole at $q^2 = 0$ becomes an $M^2$ independent constant term which we have denoted by $K^{88}$ in our sum rule analysis.

Similar considerations hold for $P^{08}_L(q^2)$. We have again taken $m_u = m_d = 0$, $m_s \neq 0$, the Ward identity

$$\lim_{q_\mu \to 0} q^{\mu}\, \Pi^{08}_{\mu\nu}(q)\, q^{\nu} = i\,\frac{-12}{\sqrt{18}} \int d^4x \langle 0|\,T\{Q_5(x), D_s(0)\}\,|0\rangle$$

$$- \frac{2}{\sqrt{18}} \int d^4x \langle 0|\,T\{D_s(x), D_s(0)\}\,|0\rangle - \frac{8}{3\sqrt{2}} m_s \langle 0|\bar{s}s|0\rangle \qquad \ldots(A27)$$

Now from Eq.(A18) we have

$$i \int d^4x \langle 0 | T\{D_s(x), Q_5(0)\} | 0 \rangle = 0 \qquad (m_u = m_d = 0) \qquad \text{....(A28)}$$

since $\chi(0) = 0$ and $D(x) = D_s(x)$ when $m_u = m_d = 0$. We can then drop the first term in the r.h.s. of Eq.(A27), in which case, neglecting higher order terms and apart from an overall factor of $-\sqrt{2}$, it is identical to the Eqs.(A23) and (A24). Thus we see that the sum rules for $P_L^{00}(q^2)$, $P_L^{08}(q^2)$ and $P_L^{88}(q^2)$ can have spurious poles and must be accounted for while extracting the coefficients of the η and η' poles.

Before concluding this appendix we make a few more observations. As pointed by Leutwyler [19], χ' (0) can be computed from Eq.(A4)

$$\chi'(0) = \frac{1}{2} F^2 \, m_{red}^2 \left\{ \frac{1}{m_u^2} + \frac{1}{m_d^2} + \frac{1}{m_s^2} \right\} + \frac{\tilde{H}_0}{6} + O(m) \qquad \text{....(A29)}$$

Numerically using the phenomenological values of quark mass ratios [19] at $F = f_\pi = 92.4$ MeV, Leutwyler estimates the first term's contribution in Eq.(A29)

$$\chi'(0) = 2.2\times10^{-3}\ \text{GeV}^2 + \frac{\tilde{H}_0}{6} \qquad \text{...(A30)}$$

This should be compared with the estimates $1.82\times10^{-3}$ GeV$^2$ [17], $1.9\times10^{-3}$ GeV$^2$ [18]as well as estimates by Ioffe and collaborators $2.3\pm0.6\times10^{-3}$ GeV$^2$ [13] and $2.0\pm0.5\times10^{-3}$ GeV$^2$[15] from proton spin sum rules depending on the method used. This suggests that $\tilde{H}_0$ term in the effective Lagrangian is indeed small. Leutwyler [19] also pointed out that since $m_u, m_d \ll m_s$, the first term in Eq.(A29) can be written as

$$\chi'(0) \simeq -\frac{1}{2} F^2 \frac{m_u^2 + m_d^2}{(m_u + m_d)^2} \, .$$

As $m_u/m_d$ is varied from 0 to 1, χ'(0) varies only by a factor of 2, so is relatively robust. This is to be contrasted with χ(0) which is very sensitive to the quark masses.

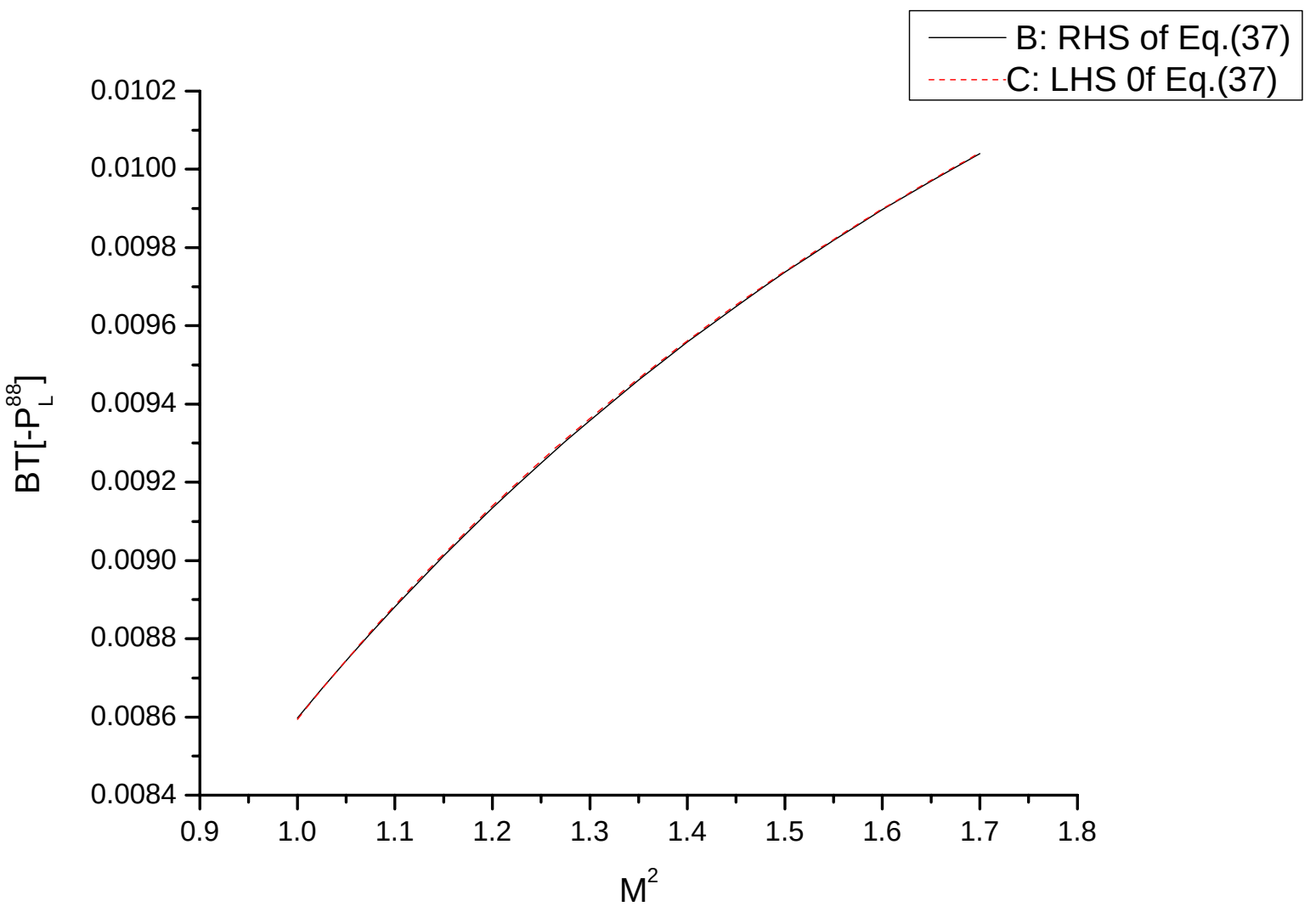

FIG.1 Plots of the two sides of the Eq.(37) (called BT[$-P_L^{88}$]), with a constant included on l.h.s, as a function of the Borel mass squared. The best fit corresponds to $K^{88} = 1.10\times10^{-3}\ GeV^4$, $m_\eta^2\left(f_\eta^8\right)^2 = 8.20\times10^{-3}\ GeV^4$ and $m_{\eta'}^2\left(f_{\eta'}^8\right)^2 = 3.55\times10^{-3}\ GeV^4$.

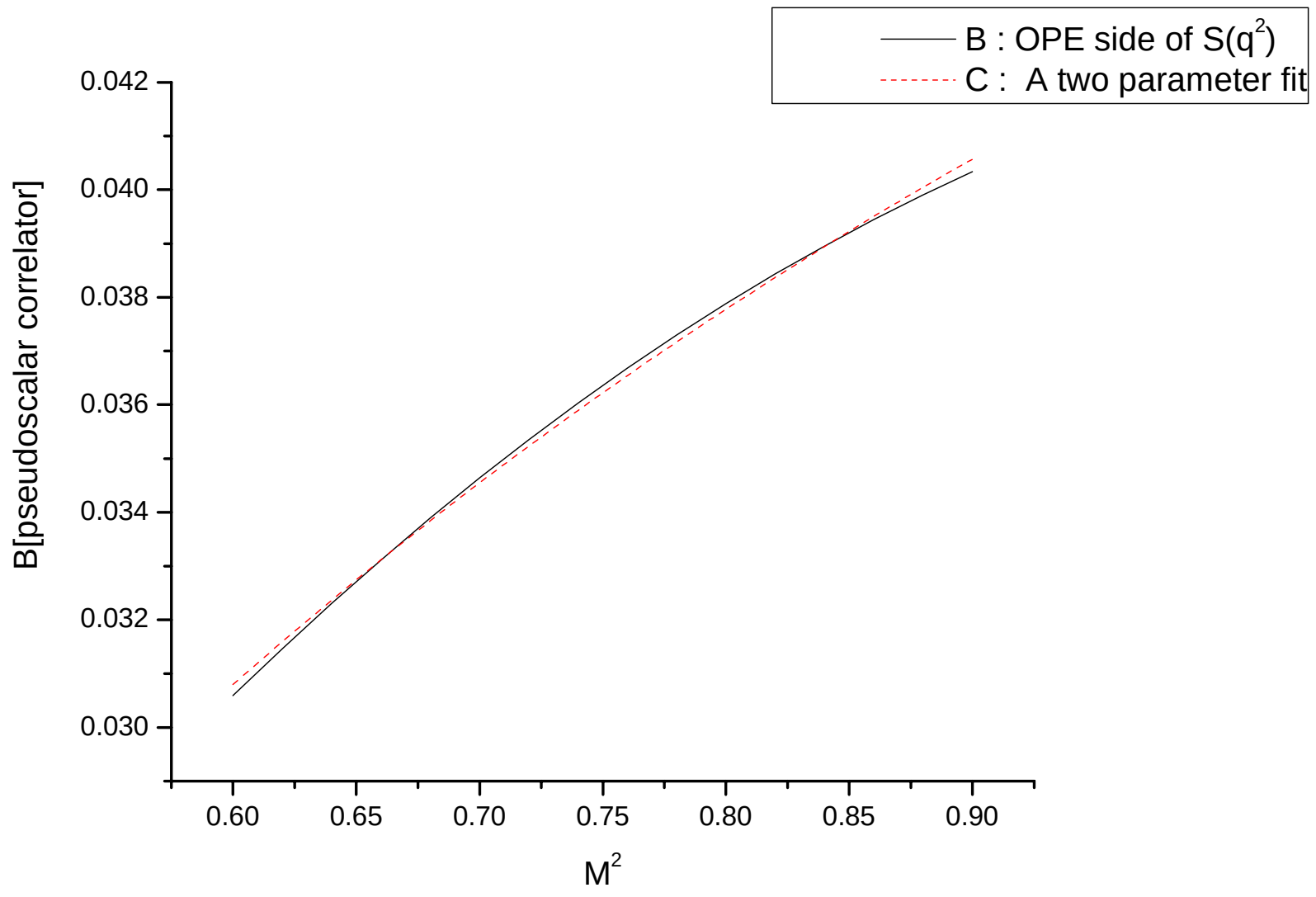

FIG. 2 The plots of Borel transforms of pseudoscalar correlator $S(q^2)$ : r.h.s. of Eq.(40) $\times \frac{3}{8m_s^2}$ (curve B) and a two-parameter fit (curve C). The fit corresponds to $3.64\times10^{-2}$ GeV$^4$ and $4.02\times10^{-2}$ GeV$^4$ as residues at η- and η'-poles. This gives $f_8$ = 168.4 MeV and $\theta_8 = \pm18.9°$.

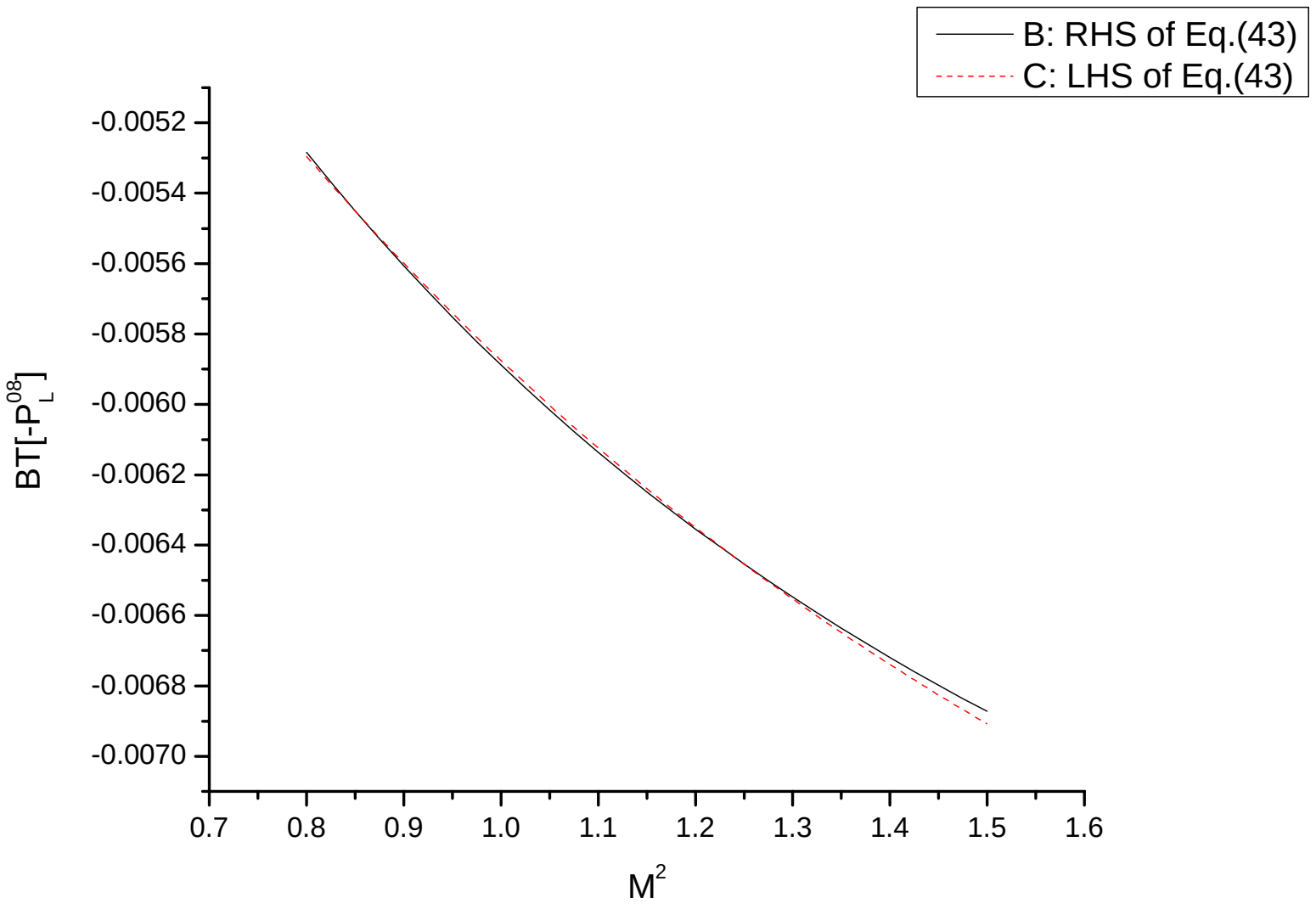

FIG. 3 Plots of the two sides of the Eq. (43) (called BT[$-P_L^{08}$]), with a constant included on l.h.s, as a function of the Borel mass squared. The best fit corresponds to $K^{08}=-3.7\times10^{-3}\ GeV^4$, $m_\eta^2 f_\eta^0 f_\eta^8=1.36\times10^{-3}\ GeV^4$ and $m_{\eta'}^2 f_{\eta'}^0 f_{\eta'}^8= -7.97\times10^{-3}\ GeV^4$.

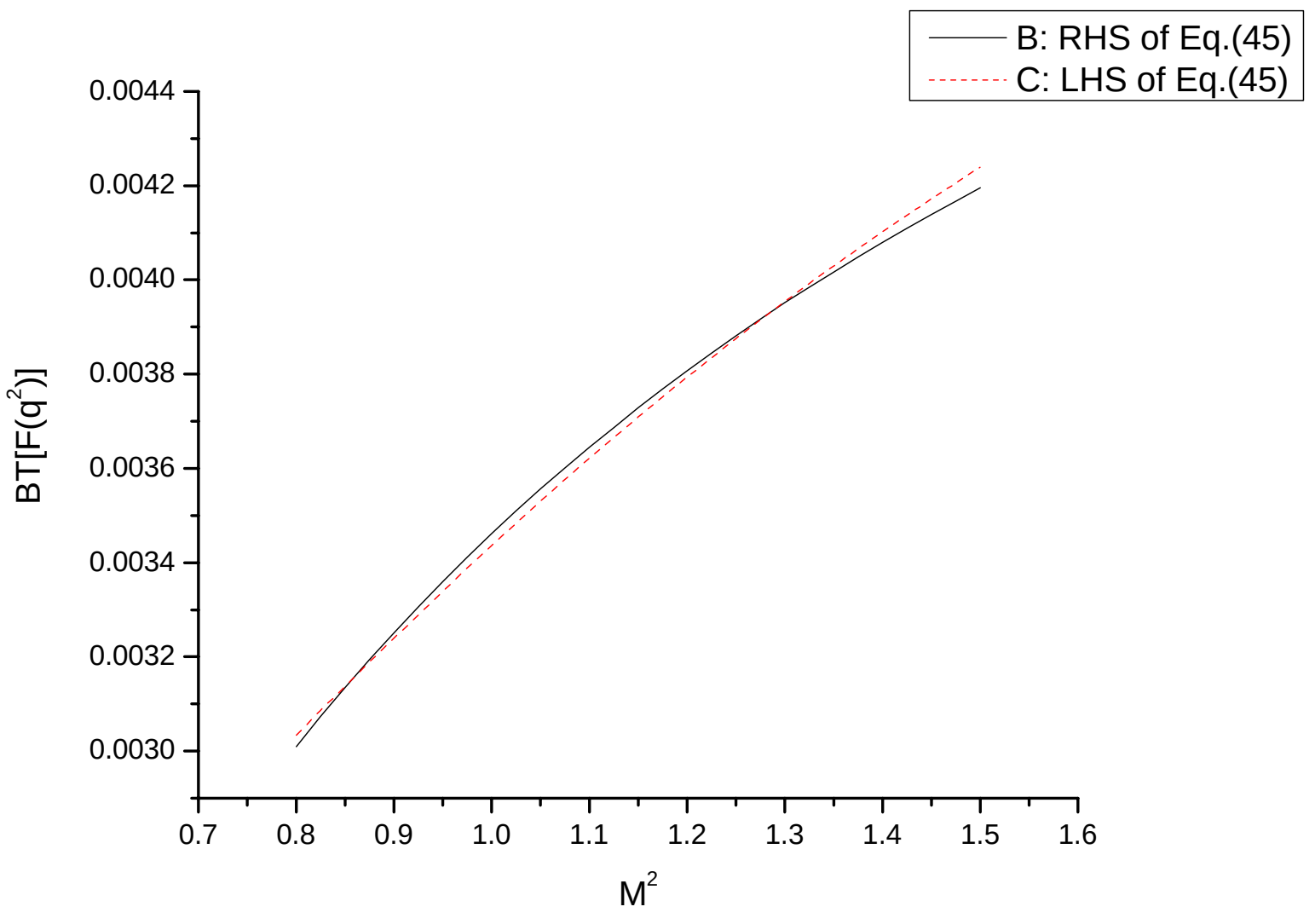

FIG. 4 Plots of BT[$-P^{00}_L(q^2)-12\chi(q^2)/q^2$] and its 3-parameter fit. The fit corresponds to a constant K=0.00464 GeV$^4$ and residues as $-0.00658$GeV$^4$ and 0.0092 GeV$^4$ and at η- and η′- poles.

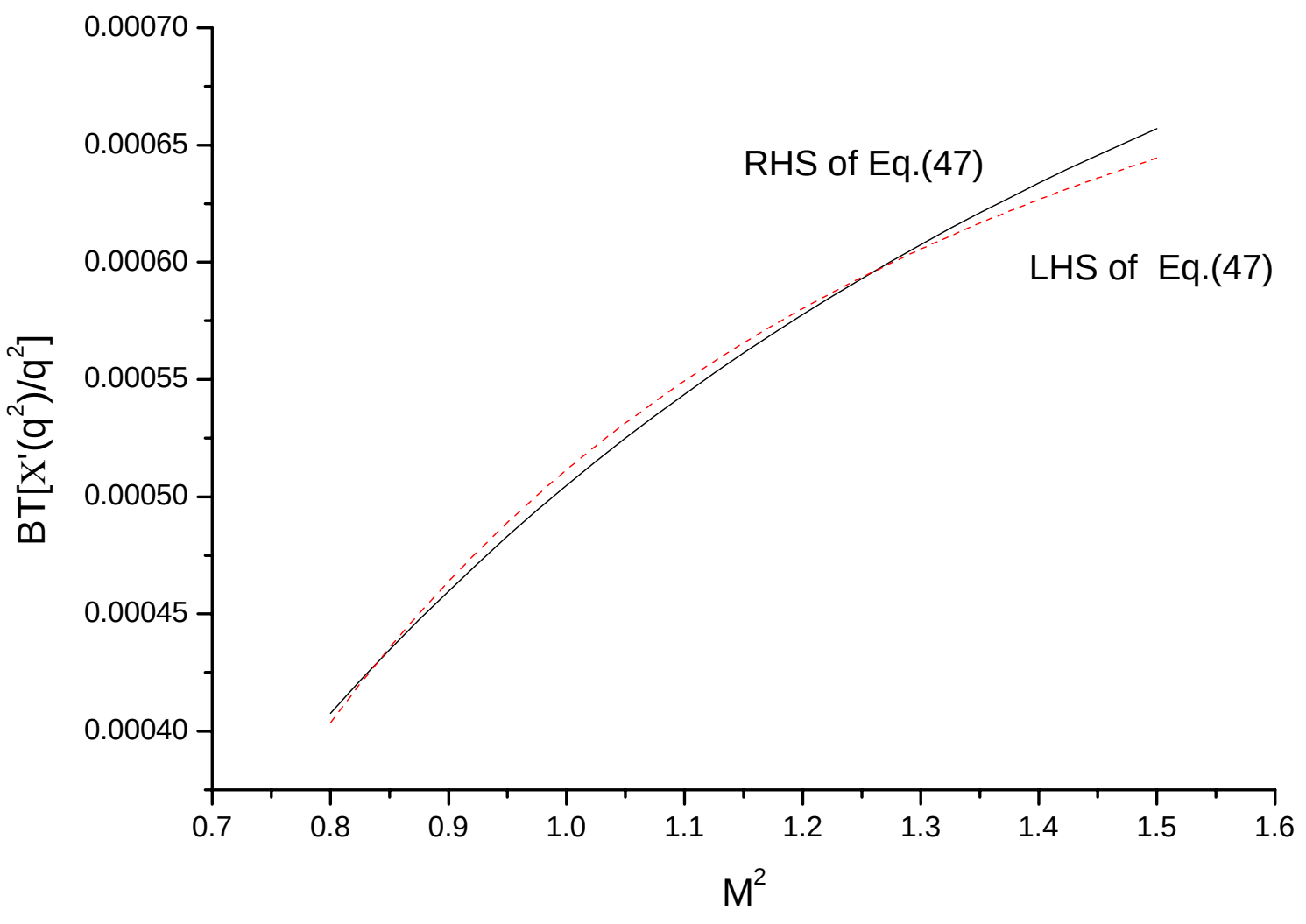

FIG. 5 Plots of BT[$\chi'(q^2)/q^2$] and its three-parameter fit as a function of Borel mass squared. The fit corresponds to 0.00165 GeV$^2$, 0.00147 GeV$^2$ and 0.000967 GeV$^2$ as the the constant [$\chi'(0)$] and the coefficients of $(1+m_\eta^2/M^2)\exp(-m_\eta^2/M^2)$ and $(1+m_{\eta'}^2/M^2)\exp(-m_{\eta'}^2/M^2)$ respectively.

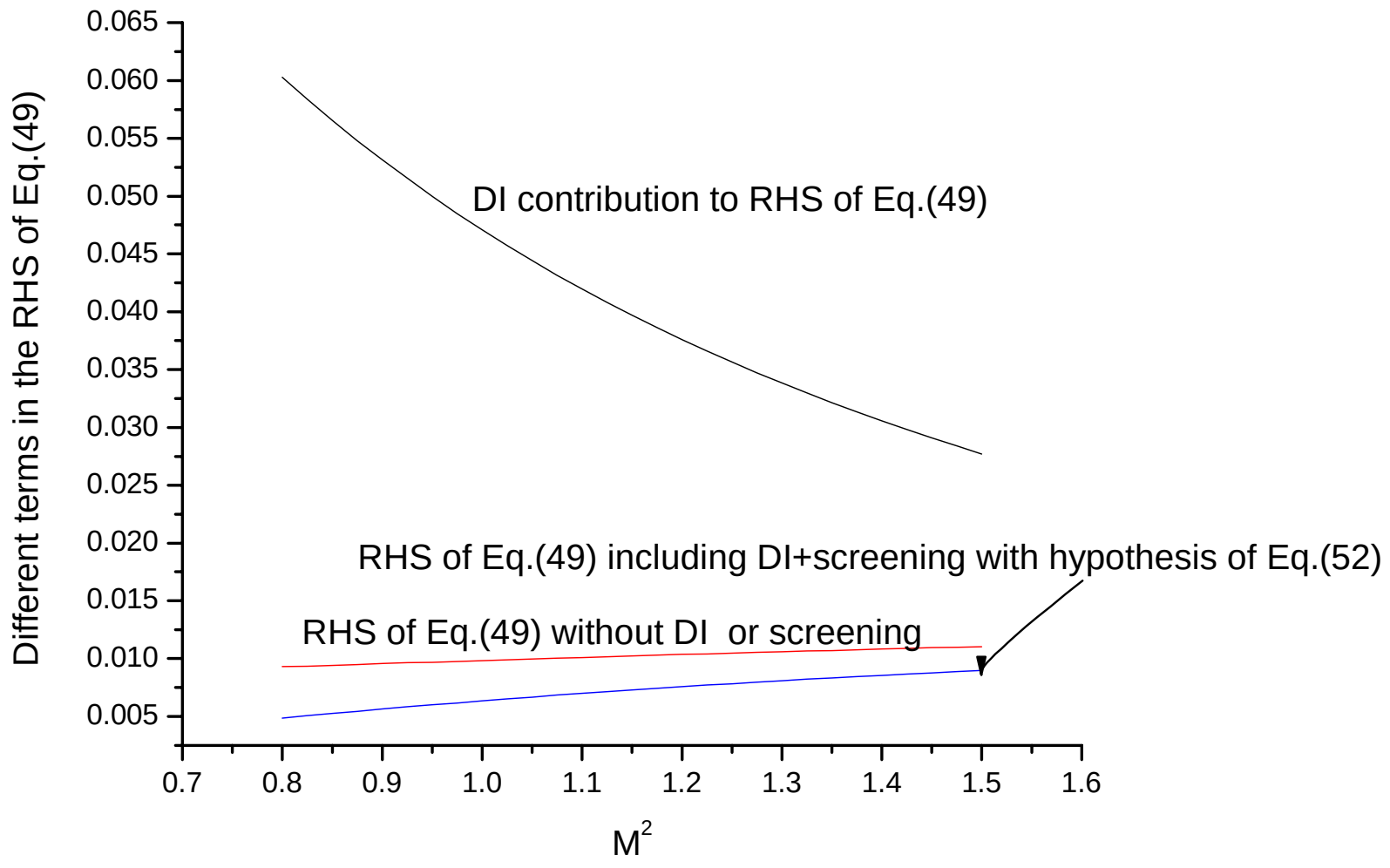

FIG. 6 Plots of direct instanton contribution (DI) to the OPE side of Eq.(49), the OPE side of Eq.(49) and a combination of the two with DI contribution included with a factor of –0.074. The last curve is separately plotted in Fig.8 also. All quantities are in GeV units. Note the difference in scale in Fig.8.

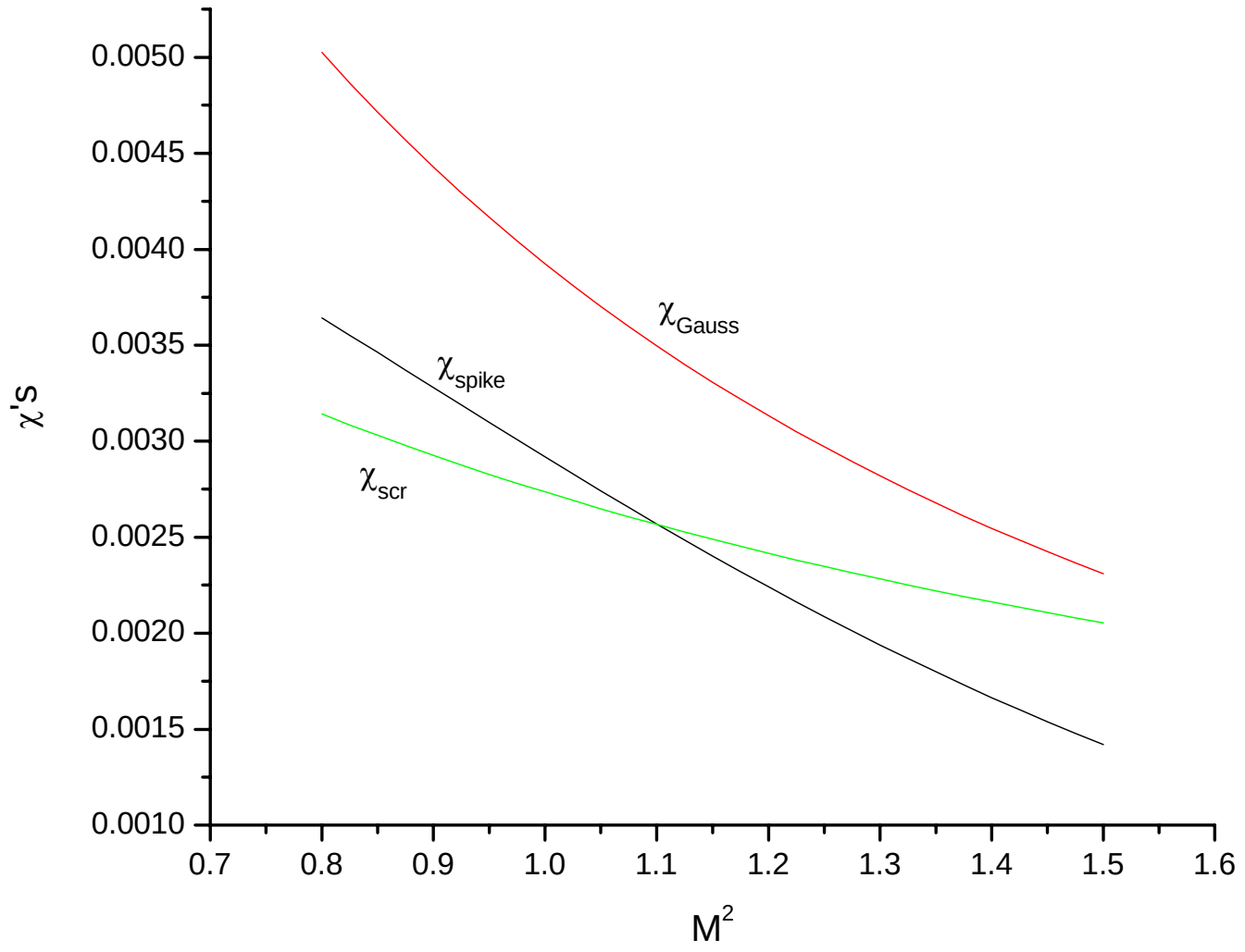

FIG.7 Plots of Borel transforms of $\chi_{DI}$ ($\chi_{Gauss}$ and $\chi_{spike}$) and $\chi_{scr}$ as a function of Borel mass parameter squared. $\chi_{scr}$ is from Forkel [21] cf. our Eqs.(50) and (51). All quantities are in GeV unit.

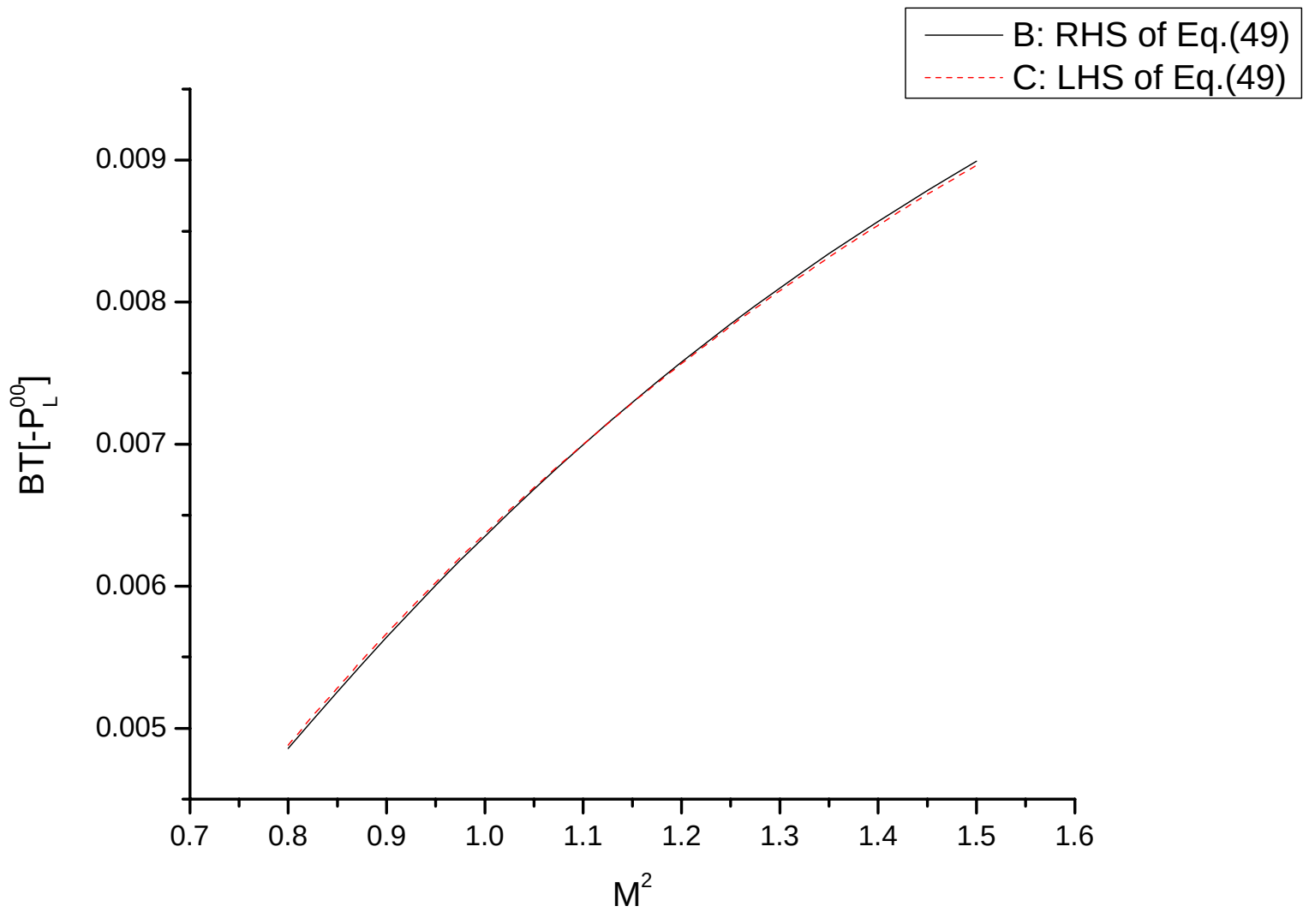

FIG. 8 Plots of two sides of Eq. (49) (called BT[$-P_L^{00}$]): OPE side with DI contribution (fraction =−0.074) included is curve B. Curve C is a three parameter fit with $K^{00}$=−9.22×$10^{-4}$ $GeV^4$ as a constant and 7.2×$10^{-5}$ $GeV^4$ and 1.812×$10^{-2}$ $GeV^4$ as residues at η- and η′- poles.

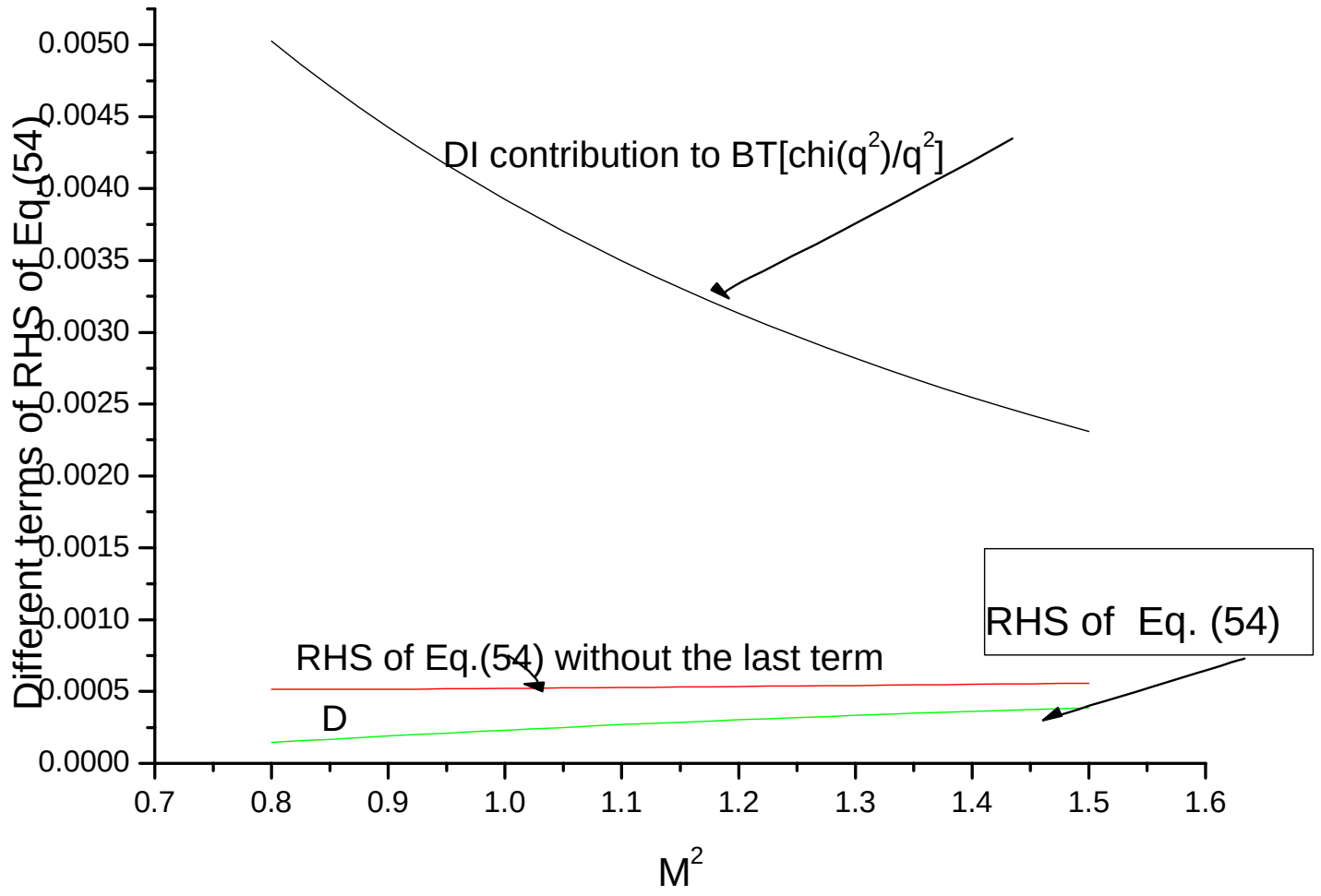

FIG. 9 Plots of BT[$\chi(q^2)/q^2$], DI contribution to BT[$\chi(q^2)/q^2$] and the combination of the two with the DI contribution appearing with a factor of -0.074 as a function of Borel mass squared. All the quantities are in GeV units. The last curve is plotted separately in Fig.10 also. Note the scale is different in Fig.10.

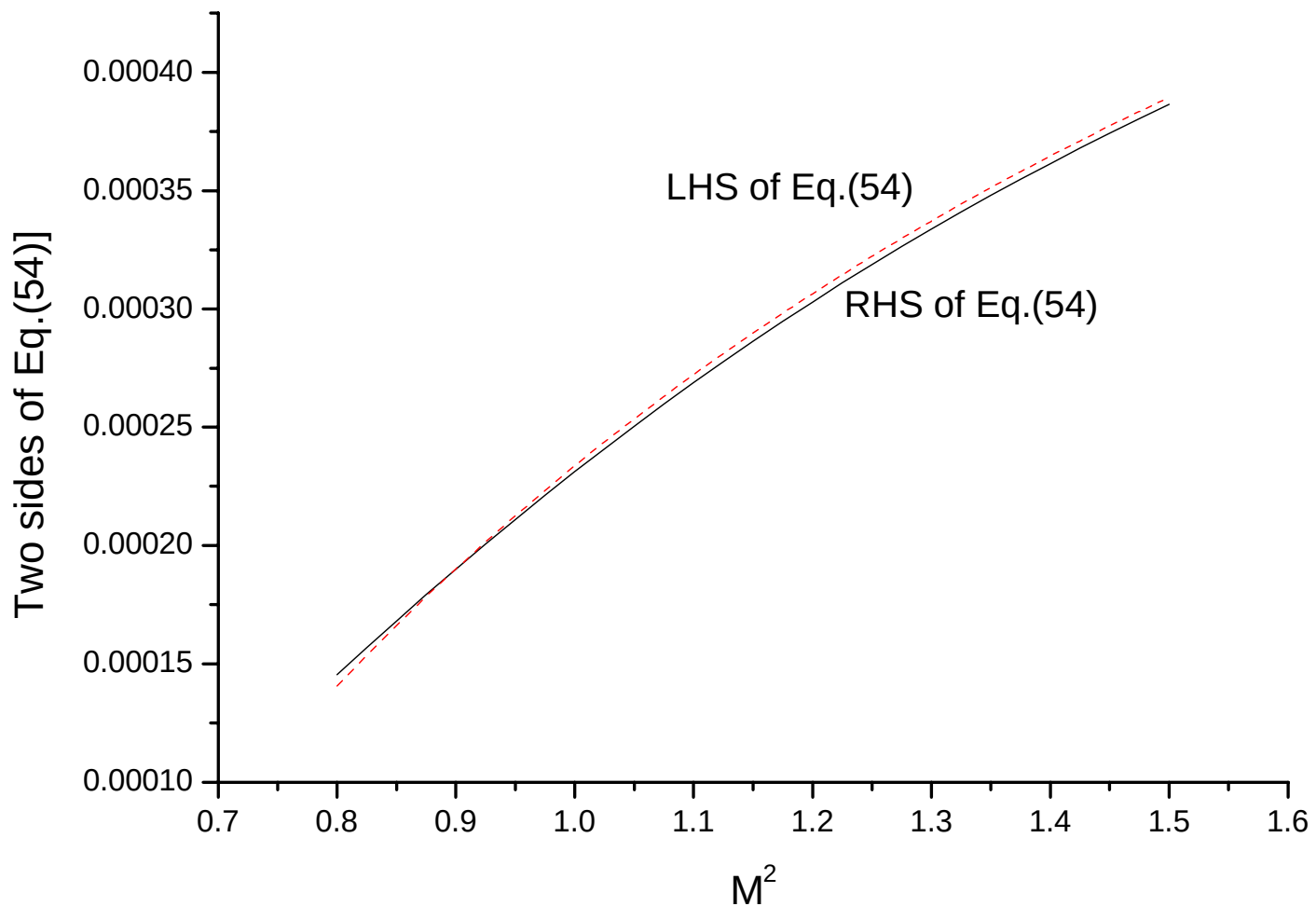

FIG. 10 Plots of BT[$\chi(q^2)/q^2$] and a three-parameter fit with a constant $-\chi(0) = -4.4\times10^{-4}$ GeV$^4$ and residues $4.4\times10^{-4}$ GeV$^4$ and $8.5\times10^{-4}$ GeV$^4$ at η- and η′- poles as a function of Borel mass squared.

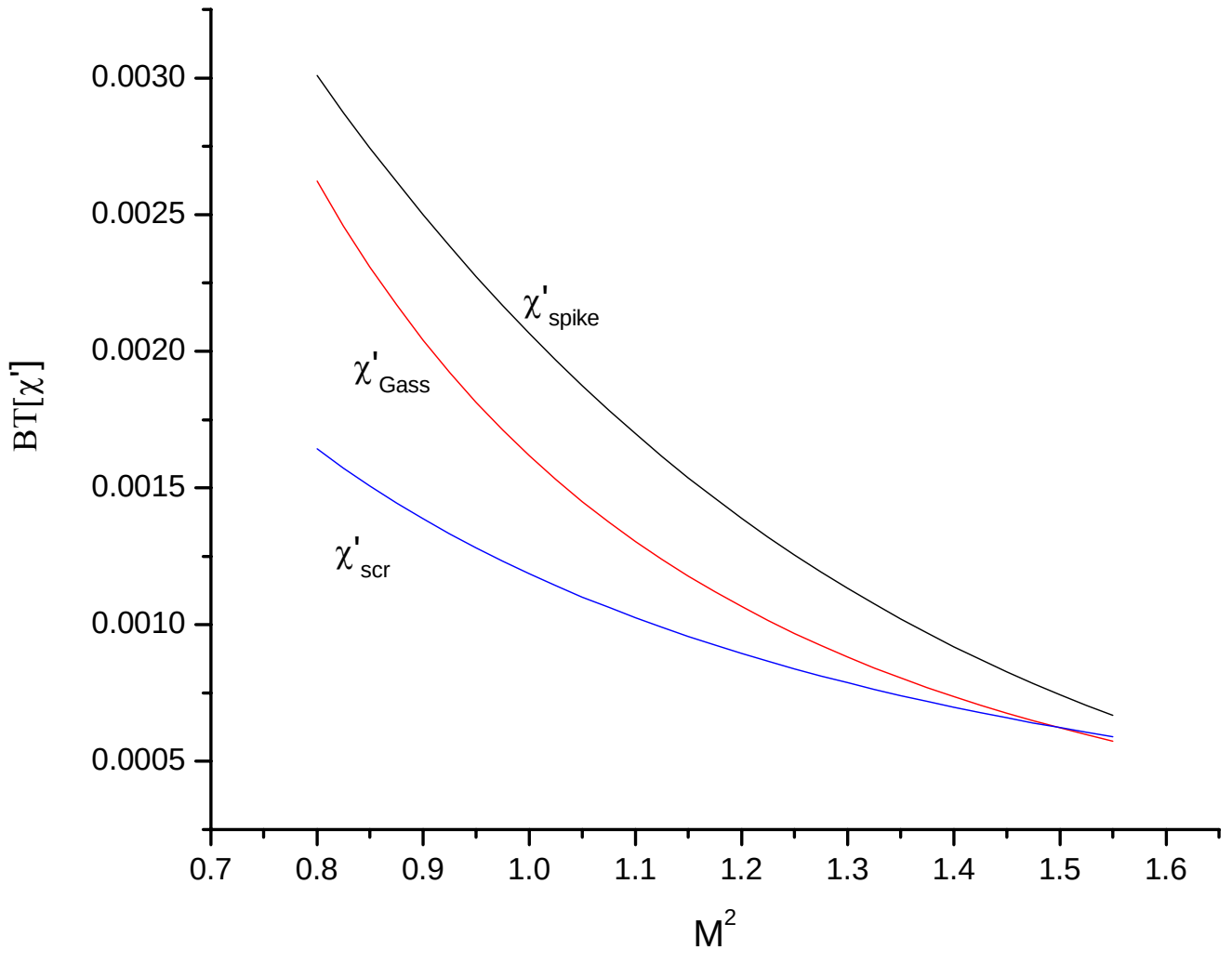

FIG. 11 Plots of BT[$\chi'_{\rm DI}$] (BT[$\chi'_{\rm Gauss}$] and BT[$\chi'_{\rm spike}$]) and BT[$\chi'_{\rm scr}$] as a function of Borel mass parameter squared ($M^2$). $\chi'_{\rm scr}$ is from Forkel [21] cf. our Eqs. (50) and (51).